\documentclass[preprint,showpacs,amsmath,amssymb,aps,prd,nofootinbib]{revtex4}
\usepackage{epsfig,graphicx,color,appendix}
\begin{document}
\begin{flushright}
KIAS-P12022
\end{flushright}
\title{\mbox{}\\[10pt]
 Non-zero $\theta_{13}$ and CP violation in a model with $A_4$ flavor symmetry}

\author{Y. H. Ahn\footnote{Email: yhahn@kias.re.kr}}

\affiliation{School of Physics, KIAS, Seoul 130-722, Korea}

\author{Sin Kyu Kang\footnote{Email: skkang@snut.ac.kr}}

\affiliation{Institute of Convergence Fundamental Studies \& \\
School of Liberal Arts, Seoul-Tech, Seoul 139-743, Korea}

\date{\today}

\begin{abstract}
Motivated by recent observations of non-zero $\theta_{13}$ from the Daya Bay and RENO experiments, we propose a renormalizable neutrino model with $A_4$
discrete symmetry accounting for deviations from the tri-bimaximal
mixing pattern of neutrino mixing matrix indicated by neutrino oscillation data.
In the model, the light neutrino masses can be generated by radiative corrections, and we show how the light neutrino mass matrix can be diagonalized
by the Pontecorvo-Maki-Nakagawa-Sakata mixing matrix whose entries are determined by the current neutrino data including the Daya Bay result.
We show that the origin of the deviations from the TBM mixing is non-degeneracy of the neutrino Yukawa coupling constants, and
unremovable CP phases in the neutrino Yukawa matrix give rise to both low energy CP violation measurable from neutrino oscillation and high energy CP violation.

\end{abstract}

\maketitle %
\section{Introduction}
Very recently, the Daya Bay Collaborations \cite{Reactor} announced $5.2\sigma$ observation of the non-zero mixing angle $\theta_{13}$ with the result given by $\sin^2 2\theta_{13}=0.092\pm 0.016(stat)\pm 0.005(syst)$
\footnote{The RENO Collaboration also announced observation of the non-zero mixing angle $\theta_{13}$ \cite{reno} in consistent with the result from the Daya Bay Collaboration.}.
This result is in good agreement with the previous data from the T2K, MINOS and Double Chooz Collaborations~\cite{Data}, and the Daya Bay and RENO progresses have led us to accomplish the measurements of three mixing angles,
$\theta_{12}, \theta_{23}$ and $\theta_{13}$ from three kinds of neutrino oscillation experiments.
A combined analysis of the data coming from T2K, MINOS, Double Chooz and Daya Bay experiments shows~\cite{Machado:2011ar} that
\begin{eqnarray}
\sin^2 2\theta_{13}=0.089\pm 0.016(0.047)~,
\end{eqnarray}
or equivalently
\begin{eqnarray}
\theta_{13}=8.68^{\circ+0.77^{\circ}~(+2.14^{\circ})}_{~-0.84^{\circ}~(-2.76^{\circ})}
 \label{expdata}
\end{eqnarray}
at $1\sigma~(3\sigma)$ levels and that the hypothesis $\theta_{13}=0$ is now rejected at a significance level higher than $6\sigma$.
In addition to the measurement of the mixing angle $\theta_{13}$, the global fit of the neutrino mixing angles and mass-squared differences at $1\sigma$ $(3\sigma)$ levels are
given by~\cite{valle}
 \begin{eqnarray}
  &&\theta_{12}=34.0^{\circ+1.0^{\circ}~(+2.9^{\circ})}_{~-0.9^{\circ}~(-2.7^{\circ})}~,
  ~\quad\theta_{23}=46.1^{\circ+3.5^{\circ}~(+7.0^{\circ})}_{~-4.0^{\circ}~(-7.5^{\circ})}~,
  ~\quad\theta_{13}=\left\{ \begin{array}{ll}
    6.5^{\circ+1.6^{\circ}~(+4.2^{\circ})}_{~-1.4^{\circ}~(-4.7^{\circ})}~, & \hbox{NH} \\
    7.3^{\circ+1.7^{\circ}~(+4.1^{\circ})}_{~-1.5^{\circ}~(-5.5^{\circ})}~, & \hbox{IH}
                      \end{array}
                    \right.
  \nonumber\\
  && \Delta m^{2}_{21}[10^{-5}{\rm eV}^{2}]=7.59^{+0.20~(+0.60)}_{-0.18~(-0.50)}~,~\qquad\Delta m^{2}_{31}[10^{-3}{\rm eV}^{2}]=\left\{\begin{array}{ll}
                2.50^{+0.09~(+0.26)}_{-0.16~(-0.36)}~, & \hbox{NH} \\
                2.40^{+0.08~(+0.27)}_{-0.09~(-0.27)}~, & \hbox{IH}
                                  \end{array}
                                \right.
 \label{expnu}
 \end{eqnarray}
in which NH and IH stand for normal hierarchical neutrino spectrum and inverted one, respectively.
The data in Eqs.~(\ref{expdata},\ref{expnu}) strongly support that the tri-bimaximal (TBM) mixing pattern of the lepton mixing matrix \cite{TBM} should be modified.
There have been theoretical attempts to explain what cause the three mixing angles to be deviated from their TBM values
\cite{dev-TB}.

Motivated by the measurements of $\theta_{13}$ from the Daya Bay and RENO experiments, we propose in this paper a renormalizable model with $A_4$ discrete symmetry which
gives rise to deviations from the TBM mixing indicated by the current neutrino data.
In addition to the leptons and the Higgs scalar of the standard model (SM), the model we porpose contains three right handed heavy Majorana neutrinos and several scalar fields
which are electroweak singlets required to construct desirable forms of the letponic mass matrices.
Although we introduce electroweak singlet heavy Majorana neutrinos, the usual seesaw mechanism does not operate
because the scalar field involved in neutrino Yukawa terms can not get vacuum expectation value (VEV).
However, as will be shown later,  the light neutrino masses can be generated through loop corrections which is a kind of
the so-called radiative seesaw mechanism \cite{Ma:2006fn}.
In the paper, we will show how the light neutrino mass matrix generated through loop corrections can be diagonalized
by the Pontecorvo-Maki-Nakagawa-Sakata (PMNS) mixing matrix whose entries are determined by the current neutrino data.
The origin of the deviations from TBM mixing in our model is non-degeneracy of the neutrino Yukawa coupling constants among three generations,
which is different from other attempts to explain the deviations from the TBM mixing \cite{dev-TB}.

Since non-trivial Dirac CP phase can exist only when the mixing angle $\theta_{13}$ has non-zero value in the standard parametrization of the leptonic mixing matrix,
the observations of non-zero $\theta_{13}$ from the Daya Bay and RENO experiments shed light on the search for CP violation in the leptonic sector.
We will show that unremovable CP phases in the neutrino Yukawa matrix are the origin of the low energy CP violation measurable from neutrino oscillation as well as high energy CP violation.
Therefore, we can anticipate that there may exist some correlation between low energy CP violation and high energy CP violation.
\section{A model with $A_4$ symmetry}

The model we consider is the standard model (SM), extended to contain three right-handed $SU(2)_{L}$-singlet Majorana neutrinos, $N_{R}$.
In addition to the usual SM Higgs doublet $\Phi$, we newly introduce two scalar fields, $\chi$ and $\eta$, that are singlet and doublet
under $SU(2)_{L}$, respectively:
 \begin{eqnarray}
  \Phi =
  \left(\varphi^{+},
  \varphi^{0}\right)^{T}~,~~~\chi~,~~~\eta =
  \left(
  \eta^{+},
  \eta^{0}
 \right)^{T}~.
  \label{Higgs}
 \end{eqnarray}
In order to account for the present neutrino oscillation data, we impose $A_{4}$ flavor symmetry for leptons and scalars.
In addition to $A_4$ symmetry, we introduce extra auxiliary $Z_{2}$ symmetry
so that a radiative seesaw at around TeV scale should operate.
Here we recall that $A_{4}$ is the symmetry group of the tetrahedron and the finite groups of the even permutation of four objects \cite{Altarelli:2005yp}.
The group $A_{4}$ has two generators $S$ and $T$, satisfying the relation $S^{2}=T^{3}=(ST)^{3}={\bf 1}$.
In the three-dimensional unitary representation, $S$ and $T$ are given by
 \begin{eqnarray}
 S={\left(\begin{array}{ccc}
 1 &  0 &  0 \\
 0 &  -1 & 0 \\
 0 &  0 &  -1
 \end{array}\right)}~,\qquad T={\left(\begin{array}{ccc}
 0 &  1 &  0 \\
 0 &  0 &  1 \\
 1 &  0 &  0
 \end{array}\right)}~.
 \label{generator}
 \end{eqnarray}
The group $A_{4}$ has four irreducible representations, one triplet ${\bf 3}$ and three singlets ${\bf 1}, {\bf 1}', {\bf 1}''$
with the multiplication rules ${\bf 3}\otimes{\bf 3}={\bf 3}_{s}\oplus{\bf 3}_{a}\oplus{\bf 1}\oplus{\bf 1}'\oplus{\bf 1}''$,
${\bf 1}'\otimes{\bf 1}''={\bf 1}$, ${\bf 1}'\otimes{\bf 1}'={\bf 1}''$
and ${\bf 1}''\otimes{\bf 1}''={\bf 1}'$.
Let's denote two $A_4$ triplets as $(a_{1}, a_{2}, a_{3})$ and $(b_{1}, b_{2}, b_{3})$,
then we have
 \begin{eqnarray}
  (a\otimes b)_{{\bf 3}_{\rm s}} &=& (a_{2}b_{3}+a_{3}b_{2}, a_{3}b_{1}+a_{1}b_{3}, a_{1}b_{2}+a_{2}b_{1})~,\nonumber\\
  (a\otimes b)_{{\bf 3}_{\rm a}} &=& (a_{2}b_{3}-a_{3}b_{2}, a_{3}b_{1}-a_{1}b_{3}, a_{1}b_{2}-a_{2}b_{1})~,\nonumber\\
  (a\otimes b)_{{\bf 1}} &=& a_{1}b_{1}+a_{2}b_{2}+a_{3}b_{3}~,\nonumber\\
  (a\otimes b)_{{\bf 1}'} &=& a_{1}b_{1}+\omega a_{2}b_{2}+\omega^{2}a_{3}b_{3}~,\nonumber\\
  (a\otimes b)_{{\bf 1}''} &=& a_{1}b_{1}+\omega^{2} a_{2}b_{2}+\omega a_{3}b_{3}~,
 \end{eqnarray}
where $\omega=e^{i2\pi/3}$ is a complex cubic-root of unity.
The representations of the field content of the model under $SU(2)\times U(1)\times A_{4}\times Z_2$ are summarized in Table-\ref{reps} :
\begin{widetext}
\begin{center}
\begin{table}[h]
\caption{\label{reps} Representations of the fields under $A_{4}\times Z_{2}$ and $SU(2)_{L}\times U(1)_{Y}$.}
\begin{ruledtabular}
\begin{tabular}{ccccccccccccc}
Field &$L_{e},L_{\mu},L_{\tau}$&$l_R,l'_R,l''_R$&$N_{R}$&$\chi$&$\Phi$&$\eta$\\
\hline
$A_4$&$\mathbf{1}$, $\mathbf{1^\prime}$, $\mathbf{1^{\prime\prime}}$&$\mathbf{1}$, $\mathbf{1^\prime}$, $\mathbf{1^{\prime\prime}}$&$\mathbf{3}$&$\mathbf{3}$&$\mathbf{1}$&$\mathbf{3}$\\
$Z_2$&$+$&$+$&$-$&$+$&$+$&$-$\\
$SU(2)_L\times U(1)_Y$&$(2,-1)$&$(1,-2)$&$(1,0)$&$(1,0)$&$(2,1)$&$(2,1)$\\
\end{tabular}
\end{ruledtabular}
\end{table}
\end{center}
\end{widetext}
With the field content and the symmetries specified in Table~\ref{reps}, the relevant renormalizable Lagrangian for the neutrino and charged lepton sectors
invariant under $SU(2)\times U(1)\times A_{4}\times Z_2$
is given by
 \begin{eqnarray}
 -{\cal L}_{\rm Yuk} &=& y^{\nu}_{1}\bar{L}_{e}(\tilde{\eta}N_{R})_{{\bf 1}}+y^{\nu}_{2}\bar{L}_{\mu}(\tilde{\eta}N_{R})_{{\bf 1}'}+y^{\nu}_{3}\bar{L}_{\tau}(\tilde{\eta}N_{R})_{{\bf 1}''}\nonumber\\
 &+&\frac{M}{2}(\overline{N^{c}_{R}}N_{R})_{{\bf 1}}+\frac{\lambda_{\chi}}{2}(\overline{N^{c}_{R}}N_{R})_{{\bf 3}_{s}} \chi\nonumber\\
 &+& y_{e}\bar{L}_{e}\Phi~l_{R}+y_{\mu}\bar{L}_{\mu}\Phi~l'_{R}+y_{\tau}\bar{L}_{\tau}\Phi~ l''_{R}
 +h.c~, \label{Lag}
 \label{lagrangian}
 \end{eqnarray}
where $\tilde{\eta}\equiv i\tau_{2}\eta^{\ast}$ with the Pauli matrix $\tau_{2}$.
Here, $L_{e,\nu,\tau}$ and $l_R^{(\prime, \prime\prime)}$ denote left handed lepton $SU(2)_L$ doublets and right handed lepton $SU(2)_L$ singlets, respectively.
The higher dimensional operators $(d\geq5)$ driven by $\chi$ and $\eta$ fields are
suppressed by a cutoff scale $\Lambda$ which is a very high energy scale.
Thus, their contributions are expected to be very small and we do not include them in this work.
In the above Lagrangian,
mass terms of the charged leptons are given by the diagonal form  because the Higgs scalar $\Phi$ and the charged lepton fields are assigned to be $A_4$ singlet.
The heavy neutrinos $N_{Ri}$ acquire a bare mass $M$ as well as a mass induced by a vacuum of electroweak singlet scalar  $\chi$ assigned to be $A_4$ triplet.
While the standard Higgs scalar $\Phi^0$ gets a VEV $v=(2\sqrt{2}G_{F})^{-1/2}=174$ GeV, the neutral component of scalar doublet $\eta$ would not acquire a nontrivial VEV
because $\eta$ has odd parity of $Z_2$ as assigned in Table~\ref{reps}
and the auxiliary $Z_{2}$ symmetry is  exactly conserved even after electroweak symmetry breaking ;
 \begin{eqnarray}
  \langle\eta^{0}_{i}\rangle=0~,~(i=1,2,3)~,\qquad\langle\Phi^{0}\rangle=\upsilon\neq0~.
 \label{vev1}
 \end{eqnarray}
 Therefore, the neutral component of scalar doublet $\eta$ can be a good dark matter candidate, and the usual seesaw mechanism does not operate because the neutrino Yukawa interactions can not generate masses.
 However, the light Majorana neutrino mass matrix can be generated radiatively through one-loop with the help of the Yukawa interaction $\bar{L}_{L}N_{R}\tilde{\eta}$ in the Lagrangian,
 which will be discussed more in detail in Sec.III.
In our model, the $A_4$ flavor symmetry is spontaneously broken by $A_4$ triplet scalars $\chi$.
From the condition of the global minima of the scalar potential, we can obtain a vacuum alignment of the fields $\chi$
relevant to achieve our goal.
%

The most general renormalizable scalar potential of $\Phi, \eta$ and $\chi$ invariant under $SU(2)_{L}\times U(1)_{Y}\times A_{4}\times Z_{2}$ is given as
 \begin{eqnarray}
  V=V(\eta)+V(\Phi)+V(\chi)+V(\eta\Phi)+V(\eta\chi)+V(\Phi\chi)
 \end{eqnarray}
where
 \begin{eqnarray}
V(\eta) &=& \mu^{2}_{\eta}(\eta^{\dag}\eta)_{\mathbf{1}}+\lambda^{\eta}_{1}(\eta^{\dag}\eta)_{\mathbf{1}}(\eta^{\dag}\eta)_{\mathbf{1}}+\lambda^{\eta}_{2}
            (\eta^{\dag}\eta)_{\mathbf{1^\prime}}(\eta^{\dag}\eta)_{\mathbf{1^{\prime\prime}}}+\lambda^{\eta}_{3}(\eta^{\dag}\eta)_{\mathbf{3}_{s}}(\eta^{\dag}\eta)_{\mathbf{3}_{s}}\nonumber\\
  &+&\lambda^{\eta}_{4}(\eta^{\dag}\eta)_{\mathbf{3}_{a}}(\eta^{\dag}\eta)_{\mathbf{3}_{a}}+\left\{\lambda^{\eta}_{5}(\eta^{\dag}\eta)_{\mathbf{3}_{s}}(\eta^{\dag}\eta)_{\mathbf{3}_{a}}+h.c.\right\}~,\nonumber\\
V(\Phi) &=& \mu^{2}_{\Phi}(\Phi^{\dag}\Phi)+\lambda^{\Phi}(\Phi^{\dag}\Phi)^{2}~,\nonumber\\
V(\chi) &=& \mu^{2}_{\chi}(\chi\chi)_{\mathbf{1}}+\lambda^{\chi}_{1}(\chi\chi)_{\mathbf{1}}(\chi\chi)_{\mathbf{1}}+\lambda^{\chi}_{2}
            (\chi\chi)_{\mathbf{1}^\prime}(\chi\chi)_{\mathbf{1}^{\prime\prime}}+\lambda^{\chi}_{3}(\chi\chi)_{\mathbf{3}_{s}}(\chi\chi)_{\mathbf{3}_{s}}\nonumber\\
  &+&\lambda^{\chi}_{4}(\chi\chi)_{\mathbf{3}_{a}}(\chi\chi)_{\mathbf{3}_{a}}
  +\lambda^{\chi}_{5}(\chi\chi)_{\mathbf{3}_{s}}(\chi\chi)_{\mathbf{3}_{a}}+\xi^{\chi}_{1}\chi(\chi\chi)_{\mathbf{3}_{s}}+\xi^{\chi}_{2}\chi(\chi\chi)_{\mathbf{3}_{a}}~,\nonumber\\
V(\eta\Phi) &=& \lambda^{\eta\Phi}_{1}(\eta^{\dag}\eta)_{\mathbf{1}}(\Phi^{\dag}\Phi)
  +\lambda^{\eta\Phi}_{2}(\eta^{\dag}\Phi)(\Phi^{\dag}\eta)+\left\{\lambda^{\eta\Phi}_{3}(\eta^{\dag}\Phi)(\eta^{\dag}\Phi)+h.c\right\}\nonumber\\
V(\eta\chi) &=& \lambda^{\eta\chi}_{1}(\eta^{\dag}\eta)_{\mathbf{1}}(\chi\chi)_{\mathbf{1}}+\lambda^{\eta\chi}_{2}(\eta^{\dag}\eta)_{\mathbf{1}^{\prime}}(\chi\chi)_{\mathbf{1}^{\prime\prime}}
  +\lambda^{\eta\chi\ast}_{2}(\eta^{\dag}\eta)_{\mathbf{1}^{\prime\prime}}(\chi\chi)_{\mathbf{1}^{\prime}}\nonumber\\
  &+&\lambda^{\eta\chi}_{3}(\eta^{\dag}\eta)_{\mathbf{3}_{s}}(\chi\chi)_{\mathbf{3}_{s}}
    (\chi\chi)_{\mathbf{3}_{s}}+\lambda^{\eta\chi}_{4}(\eta^{\dag}\eta)_{\mathbf{3}_{s}}(\chi\chi)_{\mathbf{3}_{a}}
  +\lambda^{\eta\chi}_{5}(\eta^{\dag}\eta)_{\mathbf{3}_{a}}(\chi\chi)_{\mathbf{3}_{a}}\nonumber\\
  &+&\xi^{\eta\chi}_{1}(\eta^{\dag}\eta)_{\mathbf{3}_{s}}\chi\nonumber\\
V(\Phi\chi) &=& \lambda^{\Phi\chi}(\Phi^{\dag}\Phi)(\chi\chi)_{\mathbf{1}}~.
\label{potential}
\end{eqnarray}
Here, $\mu_{\eta},\mu_{\Phi},\mu_{\chi}$, $\xi^{\chi}_{1}$, $\xi^{\chi}_{2}$, $\xi^{\eta\chi}_{1}$ and $\xi^{\eta\chi}_{2}$ have a mass dimension,
whereas $\lambda^{\eta}_{1,...,5}$, $\lambda^{\Phi}$, $\lambda^{\chi}_{1,...,5}$, $\lambda^{\eta\Phi}_{1,...,3}$, $\lambda^{\eta\chi}_{1,...,6}$ and $\lambda^{\Phi\chi}$ are all dimensionless.
In $V(\eta\Phi)$, the usual mixing term $\Phi^{\dag}\eta$ and $\Phi^{\dag}\eta\chi$ are forbidden by the $A_{4}\times Z_{2}$ symmetry.
The vacuum configuration is obtained by vanishing of the derivative of $V$ with respect to each component of the scalar fields $\Phi$ and $\chi_{i}$ but with $\langle\eta_{i}\rangle=0~(i=1,2,3)$
as follows;
 \begin{eqnarray}
  \frac{\partial V}{\partial \chi_{1}}\Big|_{<\chi_{i}>=v_{\chi_{i}}}&=&  2v_{\chi_{1}}\Big\{v^{2}_{\Phi}\lambda^{\Phi\chi}+\mu^{2}_{\chi}+(2\lambda^{\chi}_{1}-\lambda^{\chi}_{2}+4\lambda^{\chi}_{3})(v^{2}_{\chi_{2}}+v^{2}_{\chi_{3}})
  +2(\lambda^{\chi}_{1}+\lambda^{\chi}_{2})v^{2}_{\chi_{1}}\Big\}\nonumber\\
  &+&6\xi^{\chi}_{1}v_{\chi_{2}}v_{\chi_{3}}=0~,\nonumber\\
  \frac{\partial V}{\partial \chi_{2}}\Big|_{<\chi_{i}>=v_{\chi_{i}}}&=&  2v_{\chi_{2}}\Big\{v^{2}_{\Phi}\lambda^{\Phi\chi}+\mu^{2}_{\chi}+(2\lambda^{\chi}_{1}-\lambda^{\chi}_{2}+4\lambda^{\chi}_{3})(v^{2}_{\chi_{1}}+v^{2}_{\chi_{3}})
  +2(\lambda^{\chi}_{1}+\lambda^{\chi}_{2})v^{2}_{\chi_{2}}\Big\}\nonumber\\
  &+&6\xi^{\chi}_{1}v_{\chi_{1}}v_{\chi_{3}}=0~,\nonumber\\
  \frac{\partial V}{\partial \chi_{3}}\Big|_{<\chi_{i}>=v_{\chi_{i}}}&=&  2v_{\chi_{3}}\Big\{v^{2}_{\Phi}\lambda^{\Phi\chi}+\mu^{2}_{\chi}+(2\lambda^{\chi}_{1}-\lambda^{\chi}_{2}+4\lambda^{\chi}_{3})(v^{2}_{\chi_{1}}+v^{2}_{\chi_{2}})
  +2(\lambda^{\chi}_{1}+\lambda^{\chi}_{2})v^{2}_{\chi_{3}}\Big\}\nonumber\\
  &+&6\xi^{\chi}_{1}v_{\chi_{1}}v_{\chi_{2}}=0~.
 \end{eqnarray}
From those equations, we can get\footnote{There exists another nontrivial solution $\langle\chi\rangle=v_{\chi}(1,1,1)$ with $v_{\chi}=\frac{-3\xi^{\chi}_{1}\pm\sqrt{9\xi^{\chi2}_{1}-8(\mu^{2}_{\chi}+v^{2}_{\Phi}\lambda^{\Phi\chi})(3\lambda^{\chi}_{1}+4\lambda^{\chi}_{3})}}{4(3\lambda^{\chi}_{1}+4\lambda^{\chi}_{3})}$.
But, it is not desirable for our purpose.}
 \begin{eqnarray}
  \langle\chi_{1}\rangle\equiv\upsilon_{\chi}=\sqrt{\frac{-\mu^{2}_{\chi}-v^{2}_{\Phi}\lambda^{\Phi\chi}}
  {2(\lambda^{\chi}_{1}+\lambda^{\chi}_{2})}}\neq0~,\quad\langle\chi_{2}\rangle=\langle\chi_{3}\rangle=0~,
  \label{vevchi}
 \end{eqnarray}
 where $\upsilon_{\chi}$ is real.
Requiring vanishing of the derivative of $V$ with respect to $\Phi$,
 \begin{eqnarray}
  \frac{\partial V}{\partial \varphi^{0}}\Big|_{<\varphi^{0}>=v_{\Phi}}&=& 2v_{\Phi}\Big\{2v^{2}_{\Phi}\lambda^{\Phi}+\mu^{2}_{\Phi}+\lambda^{\Phi\chi}(v^{2}_{\chi1}+v^{2}_{\chi2}+v^{2}_{\chi3})\Big\}=0~,
 \end{eqnarray}
and inserting the results given by Eq.~(\ref{vevchi}), we obtain electroweak VEV,
 \begin{eqnarray}
  v\equiv v_{\Phi}=\sqrt{\frac{-\mu^{2}_{\Phi}-v^{2}_{\chi}\lambda^{\Phi\chi}}{2\lambda^{\Phi}}}~.
 \end{eqnarray}
In our scenario, we assume that $v_{\chi}$ is larger than $v_{\Phi}$.

After the breaking of the flavor and electroweak symmetries, the vacuum alignment in Eq.~(\ref{vevchi}) leads to the right-handed Majorana neutrino mass matrix
expressed as
 \begin{eqnarray}
 M_{R}=M{\left(\begin{array}{ccc}
 1 &  0 &  0 \\
 0 &  1 &  \kappa e^{i\xi} \\
 0 &  \kappa e^{i\xi} &  1
 \end{array}\right)}~,
 \label{MR1}
 \end{eqnarray}
where $\kappa=|\lambda^{s}_{\chi}\upsilon_{\chi}/M|$.
In addition, the charged lepton sector has a diagonal mass matrix $m_{\ell}=v~{\rm Diag.}(y_{e},y_{\mu},y_{\tau})$.
We note that the vacuum alignment in Eq.~(\ref{vevchi}) implies that the $A_{4}$ symmetry is spontaneously broken to its residual symmetry $Z_{2}$
in the heavy neutrino sector since $(1,0,0)$ is invariant under the generator $S$ in Eq.~(\ref{generator}).

After the scalar fields get VEVs, the Yukawa interactions in Eq.~(\ref{lagrangian}) and the charged gauge interactions in a weak eigenstate basis can be written as
 \begin{eqnarray}
 -{\cal L} &=& \frac{1}{2}\overline{N^{c}_{R}}M_{R}N_{R}+\overline{\ell_{L}}m_{\ell}\ell_{R}
 +\overline{\nu_{L}}Y_{\nu}\hat{\eta}N_{R}+\frac{g}{\sqrt{2}}W^{-}_{\mu}\overline{\ell_{L}}\gamma^{\mu}\nu_{L}+h.c~,
 \label{lagrangianA}
 \end{eqnarray}
where $\hat{\eta}={\rm Diag.}(\tilde{\eta}_{1},\tilde{\eta}_{2},\tilde{\eta}_{3})$.
One can easily see that the neutrino Yukawa matrix  is given as follows;
 \begin{eqnarray}
 Y_{\nu}=\sqrt{3}{\left(\begin{array}{ccc}
 y^{\nu}_{1} &  0 &  0 \\
 0 & y^{\nu}_{2} & 0 \\
 0 & 0 & y^{\nu}_{3}
 \end{array}\right)}U^{\dag}_{\omega}~,\qquad{\rm with}~~
 U_{\omega}=\frac{1}{\sqrt{3}}{\left(\begin{array}{ccc}
 1 &  1 &  1 \\
 1 &  \omega^{2} &  \omega \\
 1 &  \omega &  \omega^{2}
 \end{array}\right)}~.
 \label{yukawaNu}
 \end{eqnarray}
 For our convenience, let us take the basis where heavy Majorana neutrino and charged lepton mass matrices are diagonal.
 Rotating the basis
 \begin{eqnarray}
 N_{R}\rightarrow U^{\dag}_{R}N_{R}~,
 \label{basis}
 \end{eqnarray}
the right-handed Majorana mass matrix $M_{R}$ becomes real and diagonal by a unitary matrix $U_{R}$,
\begin{eqnarray}
 \hat{M}_{R}=U^{T}_{R}M_{R}U_{R}=M{\rm Diag.}(a,1,b)~,
 \label{YnuT}
 \end{eqnarray}
 where $a=\sqrt{1+\kappa^{2}+2\kappa\cos\xi}$ and $b=\sqrt{1+\kappa^{2}-2\kappa\cos\xi}$ with real and positive mass eigenvalues, $M_{1}=Ma, M_{2}=M$ and $M_{3}=Mb$.
 The unitary matrix $U_{R}$ diagonalizing $M_R$ given in Eq.(\ref{MR1}) is
\begin{eqnarray}
  U_{R} = \frac{1}{\sqrt{2}} {\left(\begin{array}{ccc}
  0  &  \sqrt{2}  &  0 \\
  1 &  0  &  -1 \\
  1 &  0  &  1
  \end{array}\right)}{\left(\begin{array}{ccc}
  e^{i\frac{\psi_1}{2}}  &  0  &  0 \\
  0  &  1  &  0 \\
  0  &  0  &  e^{i\frac{\psi_2}{2}}
  \end{array}\right)}~,
  \label{URN}
\end{eqnarray}
with the phases
\begin{eqnarray}
 \psi_1 = \tan^{-1} \Big( \frac{-\kappa\sin\xi}{1+\kappa\cos\xi} \Big)
 ~~~{\rm and}~~~ \psi_2 = \tan^{-1} \Big( \frac{\kappa\sin\xi}{1-\kappa\cos\xi} \Big)~.
\label{alphs_beta}
\end{eqnarray}
The phases $\psi_{1,2}$ go to $0$ or $\pi$ as the magnitude of $\kappa$ defined in Eq.~(\ref{MR1}) decreases.
Due to the rotation (\ref{basis}), the neutrino Yukawa matrix $Y_{\nu}$ gets modified to
 \begin{eqnarray}
 \tilde{Y}_{\nu} &=& Y_{\nu}U_{R}~, \nonumber \\
    &=& P_{\nu}^{\dag}~{\rm Diag.}(|y^{\nu}_{1}|,|y^{\nu}_{2}|,|y^{\nu}_{3}|)U^{\dag}_{\omega}U_{R}.
 \label{YnuT}
 \end{eqnarray}
Absorbing $P_{\nu}$ into the neutrino field $\nu_L$ and then transforming $\ell_{L}\rightarrow P^{\ast}_{\nu}\ell_{L}~,\quad \ell_{R}\rightarrow P^{\ast}_{\nu}\ell_{R}$,
we can make $P_{\nu}$ disappeared in $ \tilde{Y}_{\nu}$ as well as the Lagrangian Eq.(\ref{Lag}).
Then, the neutrino fields $\nu_{L}$ in the weak basis are simply transformed into the mass basis by the lepton mixing matrix,
$U_{\rm PMNS}$, so-called PMNS mixing matrix.

The lepton mixing matrix $U_{\rm PMNS}$ can be written in terms of three mixing angles and three $CP$-odd phases (one for the Dirac neutrino and two for the Majorana neutrino) as follows \cite{PDG}
 \begin{eqnarray}
  U_{\rm PMNS}={\left(\begin{array}{ccc}
   c_{13}c_{12} & c_{13}s_{12} & s_{13}e^{-i\delta_{CP}} \\
   -c_{23}s_{12}-s_{23}c_{12}s_{13}e^{i\delta_{CP}} & c_{23}c_{12}-s_{23}s_{12}s_{13}e^{i\delta_{CP}} & s_{23}c_{13}  \\
   s_{23}s_{12}-c_{23}c_{12}s_{13}e^{i\delta_{CP}} & -s_{23}c_{12}-c_{23}s_{12}s_{13}e^{i\delta_{CP}} & c_{23}c_{13}
   \end{array}\right)}Q_{\nu}~,
 \label{PMNS}
 \end{eqnarray}
where $s_{ij}\equiv \sin\theta_{ij}$ and $c_{ij}\equiv \cos\theta_{ij}$, and $Q_{\nu}={\rm Diag.}(e^{-i\varphi_{1}/2},e^{-i\varphi_{2}/2},1)$.
Here, we notice that the origin of the CP phases in $U_{\rm PMNS}$ is the CP phases $\psi_1,\psi_2$ (or $\xi$) originally coming from $M_{R}$ as can be seen by comparing Eqs.~(\ref{MR1}-\ref{YnuT}).
Thus, we expect that there can be some correlation between low energy CP violation measurable from neutrino oscillations and high energy CP violation responsible for leptogenesis in the neutrino sector.
%
%

\section{Neutrino masses and mixing angles}
\begin{figure}[t]
\begin{center}
\includegraphics*[width=0.4\textwidth]{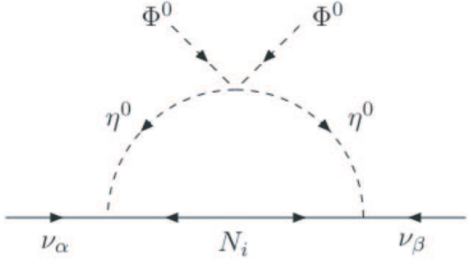}
\caption{\label{Fig1} One-loop generation of light neutrino masses.}
\end{center}
\end{figure}
We now proceed to investigate the low energy neutrino observables.
Due to the auxiliary $Z_{2}$ symmetry, the usual seesaw mechanism does not operate any more, and thus light neutrino masses can not be generated at tree level. However, similar to the scenario presented in ~\cite{Ma:2006fn}, the light neutrino mass matrix can be generated through one loop diagram drawn in Fig.~\ref{Fig1} thanks to the quartic scalar interactions. After electroweak symmetry breaking,
the light neutrino masses in the flavor basis where the charged lepton mass matrix is real and diagonal are written as
 \begin{eqnarray}
   (m_{\nu})_{\alpha\beta}=\sum_{i}\frac{\Delta m^{2}_{\eta_i}}{16\pi^{2}}\frac{(\tilde{Y}_{\nu})_{\alpha i}(\tilde{Y}_{\nu})_{\beta i}}{M_{i}}f\left(\frac{M^{2}_{i}}{\bar{m}^{2}_{\eta_i}}\right),
 \label{lownu1}
 \end{eqnarray}
where
 \begin{eqnarray}
   f(z_{i})&=&\frac{z_{i}}{1-z_{i}}\left[1+\frac{z_{i}\ln z_{i}}{1-z_{i}}\right]~,\quad \Delta m^{2}_{\eta_{i}}\equiv|m^{2}_{R_i}-m^{2}_{I_i}|=4v^{2}\lambda^{\Phi\eta}_{3}~,
 \label{lambda}
 \end{eqnarray}
with $z_{i}=M^{2}_{i}/\bar{m}^{2}_{\eta_i}$. The explicit expressions for $\bar{m}^{2}_{\eta_i}$ are presented in the Appendix.
Here, $m_{R_i}(m_{I_i})$ is the mass of the field component $\eta^{0}_{R_i}(\eta^{0}_{I_i})$ and $m^{2}_{R_i(I_i)}=\bar{m}^{2}_{\eta_i}\pm\Delta m^{2}_{\eta_i}/2$ where the subscripts $R$ and $I$ indicate real and imaginary component, respectively.
With $\tilde{M}_{R}={\rm Diag}(M_{r1},M_{r2},M_{r3})$ and $M_{ri}\equiv M_{i}f^{-1}(z_{i})$, the above formula Eq.~(\ref{lownu1}) can be expressed as
 \begin{eqnarray}
  m_{\nu} &=& \frac{v^2\lambda^{\Phi\eta}_{3}}{4\pi^{2}}\tilde{Y}_{\nu}\tilde{M}^{-1}_{R}\tilde{Y}^{T}_{\nu}
  = U_{\rm PMNS}~{\rm Diag.}(m_{1},m_{2},m_{3}) U^{T}_{\rm PMNS} \nonumber\\
  &=& m_{0}{\left(\begin{array}{ccc}
  Ay^{2}_{1} & By_{1}y_{2} & By_{1} \\
  By_{1}y_{2} & Dy^{2}_{2} & Gy_{2}  \\
  By_{1} & Gy_{2} & D
 \end{array}\right)}~,
 \label{radseesaw1}
 \end{eqnarray}
where $m_{i}(i=1,2,3)$ are the light neutrino mass eigenvalues,  $y_{1(2)}=y^{\nu}_{1(2)}/y^{\nu}_3$, and
 \begin{eqnarray} A&=&f(z_{2})+\frac{2e^{i\psi_{1}}f(z_{1})}{a}~,~~\qquad \qquad \qquad \quad B=f(z_{2})-\frac{e^{i\psi_{1}}f(z_{1})}{a}~,\nonumber\\
  D&=&f(z_{2})+\frac{e^{i\psi_{1}}f(z_{1})}{2a}-\frac{3e^{i\psi_{2}}f(z_{3})}{2b}~,\qquad
  m_{0}= \frac{v^2|y^{\nu}_{3}|^{2}\lambda^{\Phi\eta}_{3}}{4\pi^{2}M}~,\nonumber\\
  G&=&f(z_{2})+\frac{e^{i\psi_{1}}f(z_{1})}{2a}+\frac{3e^{i\psi_{2}}f(z_{3})}{2b}~.
 \label{entries}
 \end{eqnarray}
It is worthwhile to notice that in the limit of $y_{2}\rightarrow1$ the above mass matrix in Eq.~(\ref{radseesaw1}) goes to $\mu-\tau$ symmetry leading to $\theta_{13}=0$ and $\theta_{23}=-\pi/4$. Moreover, in the limit of $y_{1}, y_{2}\rightarrow1$ the above mass matrix gives TBM angles and mass eigenvalues, respectively,
 \begin{eqnarray}
  \theta_{13}&=&0, \qquad\qquad\theta_{23}=-\frac{\pi}{4}~,\qquad\qquad\theta_{12}=\sin^{-1}\left(\frac{1}{\sqrt{3}}\right)~,\nonumber\\
  m_{1}&=&3m_{0}\frac{f(z_{1})}{a}e^{i\psi_{1}}~,\quad m_{2}=3m_{0}f(z_{2})~,\quad m_{3}=3m_{0}\frac{f(z_{3})}{b}e^{i(\psi_{2}+\pi)}~,
  \end{eqnarray}
indicating that mass eigenvalues are divorced from mixing angles.
However,
recent neutrino data including the observations of non-zero $\theta_{13}$ requires deviations of $y_{1,2}$ from unit.

Now, let us show how deviations of $y_{1,2}$ from unit are responsible for non-vanishing $\theta_{13}$, and they are related with neutrino mass eigenvalues.
To separately obtain real values for the neutrino mixing angles and masses, we diagonalize the hermitian matrix $m_{\nu}m^{\dag}_{\nu}$ with $m_{\nu}$ given by Eq.~(\ref{radseesaw1}),
 \begin{eqnarray}
 m_{\nu}m^{\dag}_{\nu}&=& m^{2}_{0}\left(\begin{array}{ccc}
  \tilde{A}y^{2}_{1} & y_{1}y_{2}\left(\frac{P-Q}{2}-i\frac{3(R+S)}{2}\right) & y_{1}\left(\frac{P+Q}{2}-i\frac{3(R-S)}{2}\right) \\
  y_{1}y_{2}\left(\frac{P-Q}{2}+i\frac{3(R+S)}{2}\right) & y^{2}_{2}\frac{\tilde{F}+\tilde{G}-\tilde{K}}{4} & y_{2}\left(\frac{\tilde{F}-\tilde{G}}{4}-i\frac{3\tilde{D}}{2}\right) \\
  y_{1}\left(\frac{P+Q}{2}+i\frac{3(R-S)}{2}\right) & y_{2}\left(\frac{\tilde{F}-\tilde{G}}{4}+i\frac{3\tilde{D}}{2}\right) & \frac{\tilde{F}+\tilde{G}-\tilde{K}}{4}
  \end{array}\right)\nonumber\\
  &=& U_{\rm PMNS}~{\rm Diag.}(m^{2}_{1},m^{2}_{2},m^{2}_{3})~U^{\dag}_{\rm PMNS}~,
 \label{MM}
 \end{eqnarray}
where $\tilde{A},\tilde{D},\tilde{F},\tilde{G},\tilde{K},P,Q,R$ and $S$ are real :
 \begin{eqnarray}
  \tilde{A}&=&(1+4y^{2}_{1}+y^{2}_{2})\frac{f^{2}(z_{1})}{a^{2}}+(1+y^{2}_{1}+y^{2}_{2})f^{2}(z_{2})-2(1-2y^{2}_{1}+y^{2}_{2})\frac{f(z_{1})f(z_{2})}{a}\cos\psi_{1}~,\nonumber\\
  \tilde{F}&=& (1+4y^{2}_{1}+y^{2}_{2})\frac{f^{2}(z_{1})}{a^{2}}+4(1+y^{2}_{1}+y^{2}_{2})f^{2}(z_{2})+4(1-2y^{2}_{1}+y^{2}_{2})\frac{f(z_{1})f(z_{2})}{a}\cos\psi_{1}~,\nonumber\\
  \tilde{K}&=& 6(1-y^{2}_{2})\frac{f(z_{3})}{b}\left(\frac{f(z_{1})}{a}\cos\psi_{12}+2f(z_{2})\cos\psi_{2}\right)~,\nonumber\\
  \tilde{G}&=& 9(1+y^{2}_{2})\frac{f^{2}(z_{3})}{b^{2}}~,\nonumber\\
  \tilde{D}&=& (1-y^{2}_{2})\frac{f(z_{3})}{b}\left(\frac{f(z_{1})}{a}\sin\psi_{12}-2f(z_{2})\sin\psi_{2}\right)~,\nonumber\\
  P&=&-(1+4y^{2}_{1}+y^{2}_{2})\frac{f^{2}(z_{1})}{a^{2}}+2(1+y^{2}_{1}+y^{2}_{2})f^{2}(z_{2})-(1-2y^{2}_{1}+y^{2}_{2})\frac{f(z_{1})f(z_{2})}{a}\cos\psi_{1}~,\nonumber\\
  Q&=& 3(1-y^{2}_{2})\frac{f(z_{3})}{b}\left(\frac{f(z_{1})}{a}\cos\psi_{12}-f(z_{2})\cos\psi_{2}\right)~,\nonumber\\
  R&=& (1-2y^{2}_{1}+y^{2}_{2})\frac{f(z_{1})f(z_{2})}{a}\sin\psi_{1}~,\nonumber\\
  S&=& (1-y^{2}_{2})\frac{f(z_{3})}{b}\left(\frac{f(z_{1})}{a}\sin\psi_{12}+f(z_{2})\sin\psi_{2}\right)~,
 \label{MMele}
 \end{eqnarray}
 with $\psi_{ij}\equiv\psi_{i}-\psi_{j}$.
 To see how neutrino mass matrix given by Eq.(\ref{radseesaw1}) can lead to the deviations of neutrino mixing angles from their TBM values,
 we first introduce three small quantities $\epsilon_{i},~(i=1-3)$ which are responsible for the deviations of the $\theta_{jk}$ from their TBM values ;
 \begin{eqnarray}
  \theta_{23}=-\frac{\pi}{4}+\epsilon_{1}~, \qquad\theta_{13}=\epsilon_{2}~, \qquad\theta_{12}= \sin^{-1}\left(\frac{1}{\sqrt{3}}\right)+\epsilon_{3}~.
 \end{eqnarray}
Then, the PMNS mixing matrix keeping unitarity up to order of $\epsilon_{i}$ can be written as
 \begin{eqnarray}
 U_{\rm PMNS}&=&{\left(\begin{array}{ccc}
 \frac{\sqrt{2}-\epsilon_{3}}{\sqrt{3}} &  \frac{1+\epsilon_{3}\sqrt{2}}{\sqrt{3}} &  \epsilon_{2}e^{-i\delta_{CP}} \\
 -\frac{1+\epsilon_{1}+\epsilon_{3}\sqrt{2}}{\sqrt{6}}+\frac{\epsilon_{2}e^{i\delta_{CP}}}{\sqrt{3}} &  \frac{\sqrt{2}+\epsilon_{1}\sqrt{2}-\epsilon_{3}}{\sqrt{6}}+\frac{\epsilon_{2}e^{i\delta_{CP}} }{\sqrt{6}} &  \frac{-1+\epsilon_{1}}{\sqrt{2}} \\
 \frac{-1+\epsilon_{1}+\epsilon_{3}\sqrt{2}}{\sqrt{6}}-\frac{\epsilon_{2}}{\sqrt{3}}e^{i\delta_{CP}} &  \frac{\sqrt{2}-\epsilon_{3}-\sqrt{2}\epsilon_{1}}{\sqrt{6}}-\frac{\epsilon_{2}}{\sqrt{6}}e^{i\delta_{CP}} &  \frac{1+\epsilon_{1}}{\sqrt{2}}
 \end{array}\right)}Q_{\nu}
 +{\cal O}(\epsilon^{2}_{i})~.
 \label{Unu}
 \end{eqnarray}
The small deviation $\epsilon_{1}$ from maximality of atmospheric mixing angle is expressed in terms of the parameters in Eq.~(\ref{MMele}) as
 \begin{eqnarray}
  \tan\epsilon_{1}=\frac{R(1+y_{2})-S(1-y_{2})}{R(1-y_{2})-S(1+y_{2})}~.
 \label{Atmdevi}
 \end{eqnarray}
The reactor angle $\theta_{13}$ and Dirac-CP phase $\delta_{CP}$ are expressed as
 \begin{eqnarray}
  \tan2\theta_{13}&\simeq&\frac{y_{1}|\Omega|}{\sqrt{2}(\Theta-\tilde{A})}~,\nonumber\\
  \tan\delta_{CP}&=&3\frac{(R-S)^{2}+y^{2}_{2}(R+S)^{2}}{(P+Q)(R-S)-y^{2}_{2}(P-Q)(R+S)}~,
 \label{DiracCP}
 \end{eqnarray}
where
 \begin{eqnarray}
  \Omega&=&(1-y_{2})P+(1+y_{2})Q+\epsilon_{1}\{(1+y_{2})P+(1-y_{2})Q\nonumber\\
  &-&3i\Big\{R(1-y_{2})-S(1+y_{2})+\epsilon_{1}\left(R(1+y_{2})-S(1-y_{2})\right)\Big\}~,\nonumber\\
  \Theta &=&\frac{1}{4}\left\{(\tilde{F}+\tilde{G}-\tilde{K})\left(\frac{1+y^{2}_{2}}{2}+\epsilon_{1}(1-y^{2}_{2})\right)-y_{2}(\tilde{F}-\tilde{G})\right\}~.
 \label{DiracCP}
 \end{eqnarray}
In the limit of  $y_{1},y_{2}\rightarrow1$, the parameters $Q,R,S,\epsilon_{1}$ go to zero, which in turn leads to  $\theta_{13}\rightarrow0$ and $\delta_{CP}\rightarrow0$ as expected.
Finally, the solar mixing angle is given as
 \begin{eqnarray}
  \tan2\theta_{12}\simeq\frac{y_{1}Z}{\sqrt{2}(\Psi_{2}-\Psi_{1})}~,
 \label{sol1}
 \end{eqnarray}
where the parameters $\Psi_{1},\Psi_{2}$ and $Z$ with $|\epsilon_{i}|\ll1$ are given as
 \begin{eqnarray}
  \Psi_{1}&\simeq&\tilde{A}-\frac{\epsilon_{2}|\Omega|}{\sqrt{2}}~,\qquad Z\simeq P(1+y_{2})+Q(1-y_{2})-\epsilon_{1}\left\{P(1-y_{2})+Q(1+y_{2})\right\}~,\nonumber\\
  \Psi_{2}&\simeq&\frac{\tilde{F}+\tilde{G}-\tilde{K}}{8}(1+y^{2}_{2})+\frac{\tilde{F}-\tilde{G}}{4}y_{2}-\epsilon_{1}\frac{\tilde{F}+\tilde{G}-\tilde{K}}{4}(1-y^{2}_{2})~.
 \label{para1}
 \end{eqnarray}
Note that in Eq.~(\ref{sol1}) the condition $P(1+y_{2})+Q(1-y_{2})\gg|\epsilon_{1}\left\{P(1-y_{2})+Q(1+y_{2})\right\}|$ should be satisfied,
in order for the solar mixing angle $\theta_{12}$ to be lie in the allowed region from the experimental data given in Eq.~(\ref{expnu}).
The squared-mass eigenvalues of three light neutrinos are given by
 \begin{eqnarray}
    m^{2}_{1}&\simeq&m^{2}_{0}\left\{c^{2}_{12}\Psi_{1}+s^{2}_{12}\Psi_{2}-\frac{y_{1}Z}{2\sqrt{2}}\sin2\theta_{12}\right\}~,\nonumber\\
    m^{2}_{2}&\simeq&m^{2}_{0}\left\{s^{2}_{12}\Psi_{1}+c^{2}_{12}\Psi_{2}+\frac{y_{1}Z}{2\sqrt{2}}\sin2\theta_{12}\right\}~,\nonumber\\
    m^{2}_{3}&\simeq&m^{2}_{0}\left\{\Theta+\frac{\epsilon_{2}|\Omega|}{\sqrt{2}}\right\}~.
 \label{eigenvalueGen}
 \end{eqnarray}
We see from Eq.~(\ref{para1}) that the deviation $\epsilon_{3}$ from tri-maximality of solar mixing angle is roughly expressed as
 \begin{eqnarray}
  \sin\epsilon_{3}&\simeq&\frac{y_{1}3\sqrt{2}Zm^{2}_{0}}{2\Delta m^{2}_{21}}-2\sqrt{2}~.
 \end{eqnarray}

In the limit of $|\epsilon_{i}|\ll1$, the solar and atmospheric mass-squared differences are roughly given in a good approximation by
\begin{eqnarray}
 \Delta m^{2}_{\rm Sol}\equiv m^{2}_{2}-m^{2}_{1}
 &\simeq& \frac{m^{2}_{0}}{24}\Big\{(\tilde{F}+\tilde{G}-\tilde{K})(1+y^{2}_{2})+2y_{2}(\tilde{F}-\tilde{G})-8\tilde{A}\nonumber\\
 &+&16y_{1}\Big(P(1+y_{2})+Q(1-y_{2})\Big)\Big\}~,\nonumber\\
 \Delta m^{2}_{\rm Atm}\equiv m^{2}_{3}-m^{2}_{1}&\simeq&  \frac{m^{2}_{0}}{3}\Big\{\frac{\tilde{F}+\tilde{G}-\tilde{K}}{4}(1+y^{2}_{2})-y_{2}(\tilde{F}-\tilde{G})-2\tilde{A}\nonumber\\
 &-&y_{1}\Big(P(1+y_{2})+Q(1-y_{2})\Big)\Big\}~.
 \label{deltam2}
\end{eqnarray}
Here we note that the parameter $M_{ri}$ in Eq.~(\ref{radseesaw1}) can be simplified in the following limiting cases as
 \begin{eqnarray}
  M_{ri}\simeq\left\{
  \begin{array}{ll}
    M_{i}\left[\ln z_{i}-1\right]^{-1}, & \hbox{for $z_{i}\gg1$} \\
    2M_{i}, & \hbox{for $z_{i}\rightarrow1$} \\
    \bar{m}^{2}_{\eta}M^{-1}_{i}, & \hbox{for $z_{i}\ll1$~.}
  \end{array}
\right.
 \end{eqnarray}
\begin{figure}[t]
\begin{minipage}[t]{6.0cm}
\epsfig{figure=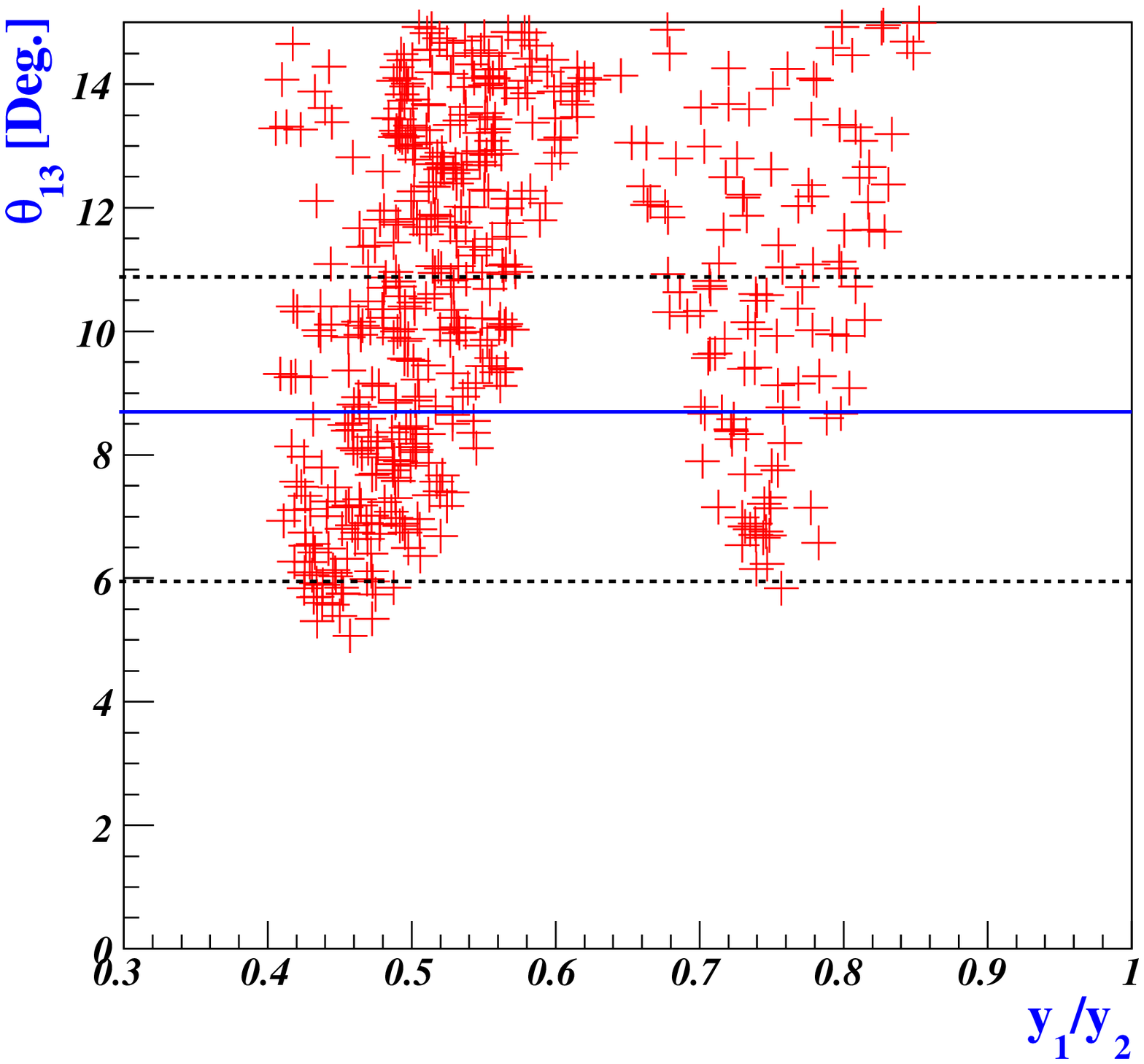,width=6.5cm,angle=0}
\end{minipage}
\hspace*{1.0cm}
\begin{minipage}[t]{6.0cm}
\epsfig{figure=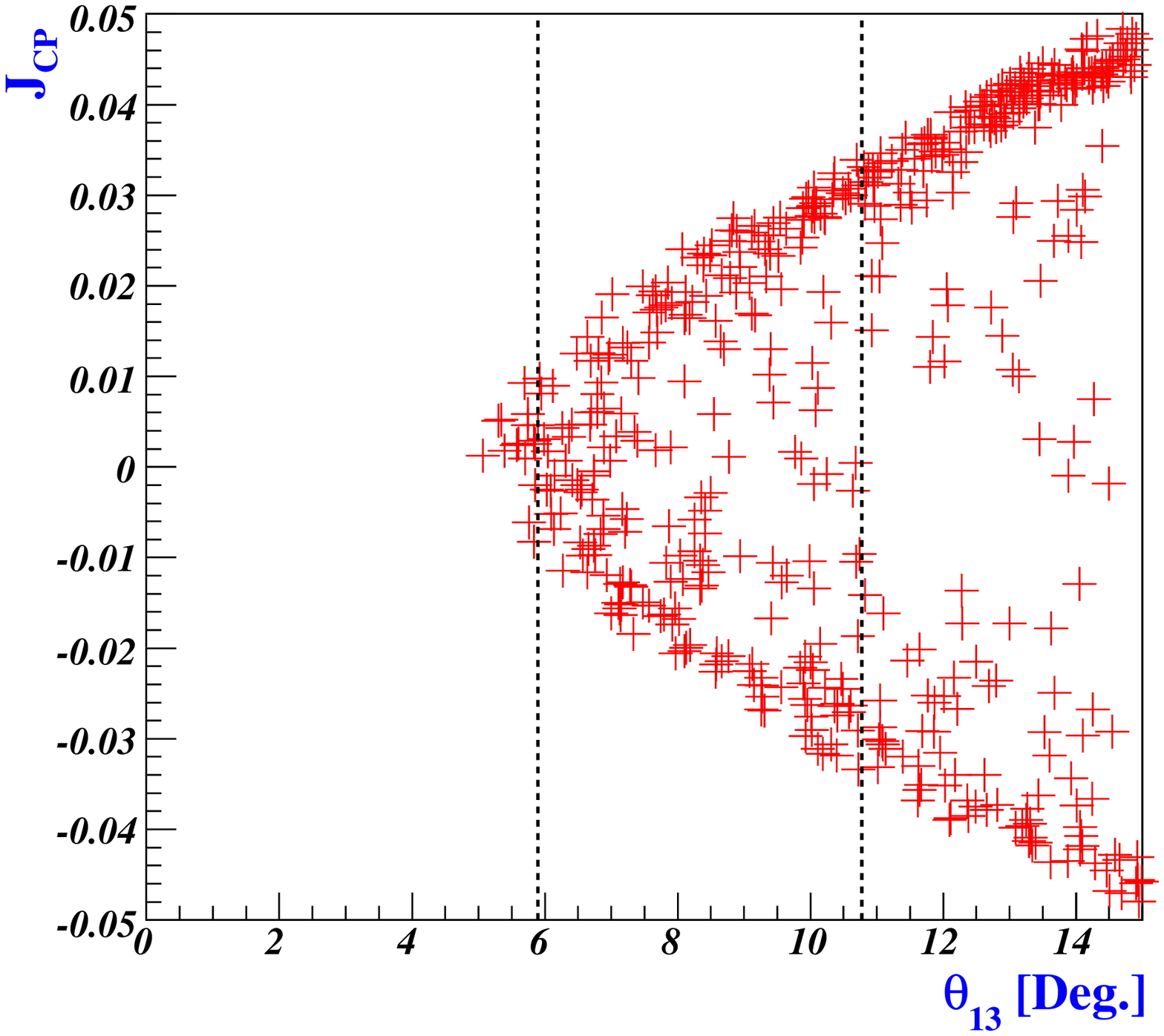,width=6.5cm,angle=0}
\end{minipage}\\
\begin{minipage}[t]{6.0cm}
\epsfig{figure=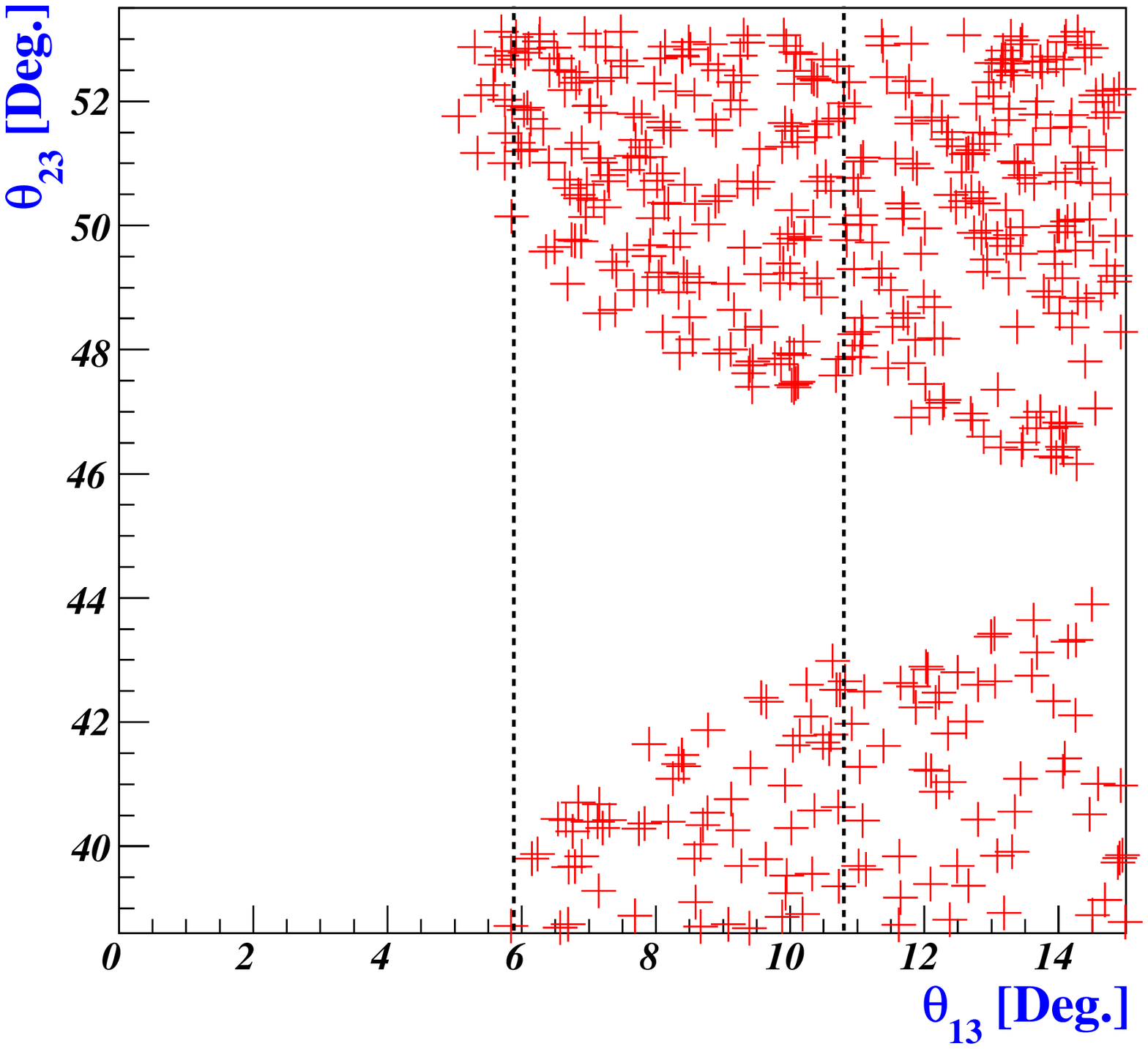,width=6.5cm,angle=0}
\end{minipage}
\hspace*{1.0cm}
\begin{minipage}[t]{6.0cm}
\epsfig{figure=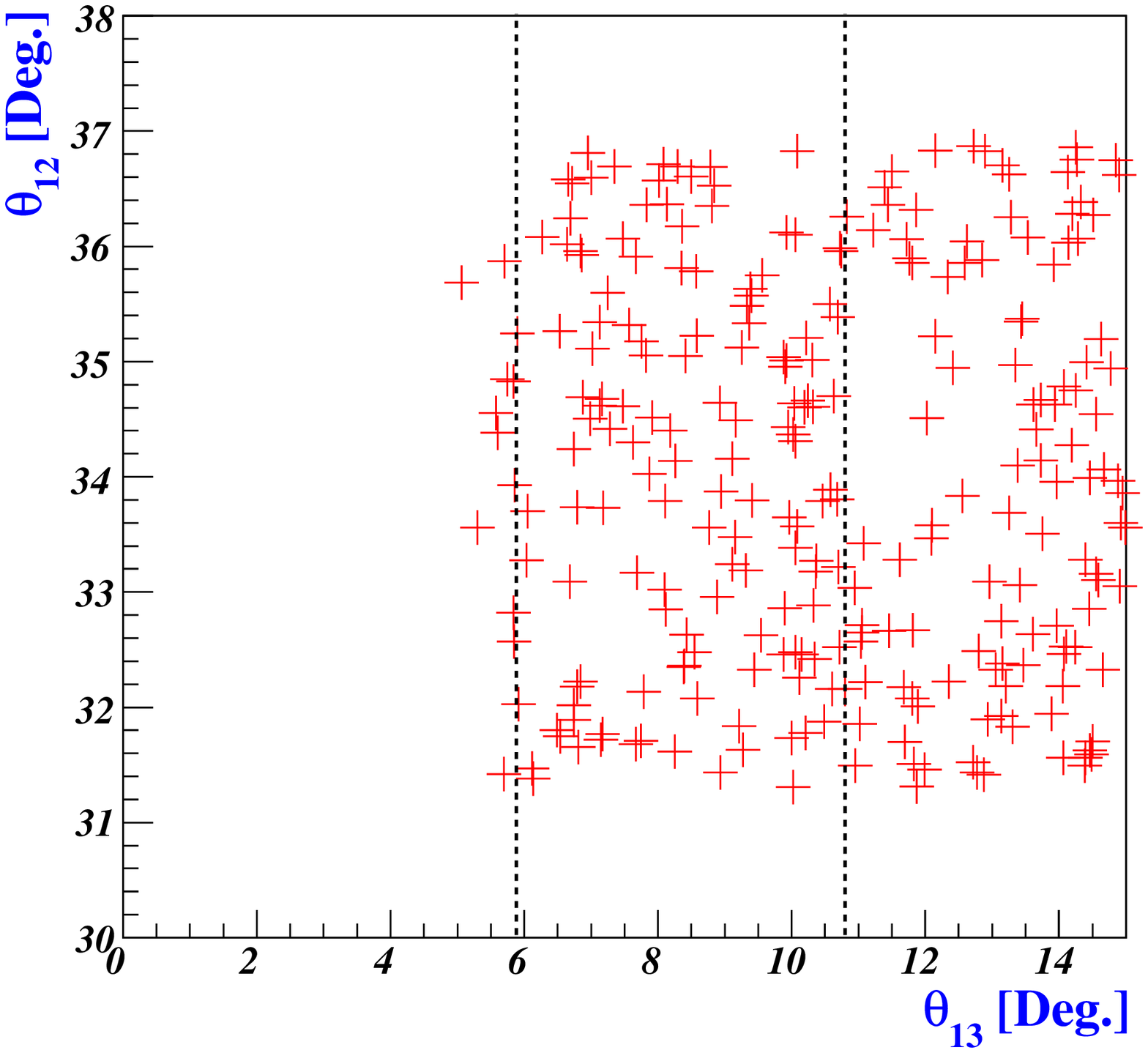,width=6.5cm,angle=0}
\end{minipage}
\caption{\label{FigA1}
Plots for Case (i) displaying the reactor mixing angle $\theta_{13}$
versus the ratio $y_{1}/y_{2}$ (upper left panel), and the Jarlskog invariant $J_{CP}$ versus the reactor angle $\theta_{13}$ (upper right panel). Allowed values for the atmospheric mixing angle $\theta_{23}$ (lower left panel) and the solar mixing angle $\theta_{12}$ (lower right panel) versus the mixing angle $\theta_{13}$, respectively.
The thick line corresponds to $\theta_{13} = 8.68^{\circ}$ which is the best-fit value of Eq.~(\ref{expdata}) including the Daya Bay result. And the horizontal and vertical dotted lines in both plots indicate the upper and lower bounds on $\theta_{13}$ given in Eq.~(\ref{expdata}) at $3\sigma$}
\end{figure}

\section{Numerical results}
As is well known, the observed hierarchy $|\Delta m^{2}_{\rm Atm}|\gg\Delta m^{2}_{\rm Sol}>0$ leads to two possible neutrino mass spectrum: (i) $m_{1}<m_{2}<m_{3}$ (normal mass spectrum), and (ii) $m_{3}<m_{1}<m_{2}$ (inverted mass spectrum).
Since there are many unknown parameters such as masses of heavy Majorana neutrinos and scalar fields $\eta_R, \eta_I$, we consider a particular parameter set for those parameters and
show how the measured values of the mixing angle $\theta_{13}$ can be accommodated in our model while keeping the other neutrino parameters such as solar and atmospheric mixing angles
and mass-squared differences are satisfied with the current data.

The mass matrix in Eq.~(\ref{radseesaw1}) contains 10 free parameters : $\lambda^{\Phi\eta}_{3},M,y^{\nu}_{3}$, $z_{1},z_{2},z_{3}$ and $y_{1},y_{2},\xi,\kappa$.
The combination of the first three of them, $\{\lambda^{\Phi\eta}_{3},M,y^{\nu}_{3}\}$, leads to the overall neutrino scale parameter $m_{0}$.
As shown above, the elements of the mass matrix in Eq.~(\ref{radseesaw1}) are expressed in terms of measurable neutrino parameters,
$\theta_{12},\theta_{13},\theta_{23},m_{1,2,3},\delta_{CP},\varphi_{1,2}$.
Among them, three mixing angles and two mass squared differences are measured.
For numerical analysis~\cite{Antusch:2005gp}, we need to fix some parameters by hand since there are too many model parameters to be predicted.
As an example, we take a case $M^2_1=\bar{m}^2_{\eta_1}$, $M^2_2=1.3\bar{m}^2_{\eta_2}$ $M^2_3=1.5\bar{m}^2_{\eta_3}$, and fix the overall seesaw scale $M$ to be $1$ TeV.
Then, the parameters $m_0,y_{1},y_{2},\kappa,\xi$ can be determined from the experimental results of three mixing angles and two mass squared differences.
In addition, the CP phases $\delta_{CP}, \varphi_{1,2}$ can be predicted after determining the model parameters.
Depending on the values of the model parameters, there exist two possibilities for the light neutrino spectrum, one is normal mass hierarchy and the other is inverted hierarchy.
In the following, we discuss the two cases separately.
\vskip 0.4cm
{\bf (i)} normal hierarchy of light neutrino
\vskip 0.5cm
Based on the formulae for the neutrino mixing angles and masses, we numerically scan the parameters $m_0, y_1, y_2, \kappa, \xi$
and then pick up the values of those five parameters which are consistent with the experimental data given at $3\sigma$  in Eq.~(\ref{expnu}). For the mixing angle $\theta_{13}$,  we a bit widely allow its value from $5^{\circ}$ to $15^{\circ}$
instead of  its experimental values at $3\sigma$\footnote{Note that very small mixing angle $\theta_{13}$ less then $1^{\circ}$ can be achieved in the case that $y_{1}\rightarrow1$ or $\sin\psi_{1}\rightarrow0$ converges more faster than $y_{2}\rightarrow1$.} .
In such a way, we can obtain the allowed regions of the parameters given by
 \begin{eqnarray}
  &&1.40<\kappa<2.38~,~~~0.44<y_{1}<0.89~,\qquad0.60<y_{2}<0.84~{~\rm and}~1.1<y_{2}<1.89~,\nonumber\\
  &&190^{\circ}\leq\xi<211^{\circ}~,\qquad\qquad 0.23\leq \frac{y^{\nu}_{3} \lambda_{3}^{\Phi\eta}}{10^{-9}}<0.46~.
  \label{input}
 \end{eqnarray}
 We found that normal mass ordering of light neutrino can be achieved when $M_{1}\lesssim M_{2}<M_{3}$ or $M_{2}\lesssim M_{1}<M_{3}$ are satisfied for the parameter spaces given above.
\begin{figure}[t]
\begin{minipage}[t]{6.0cm}
\epsfig{figure=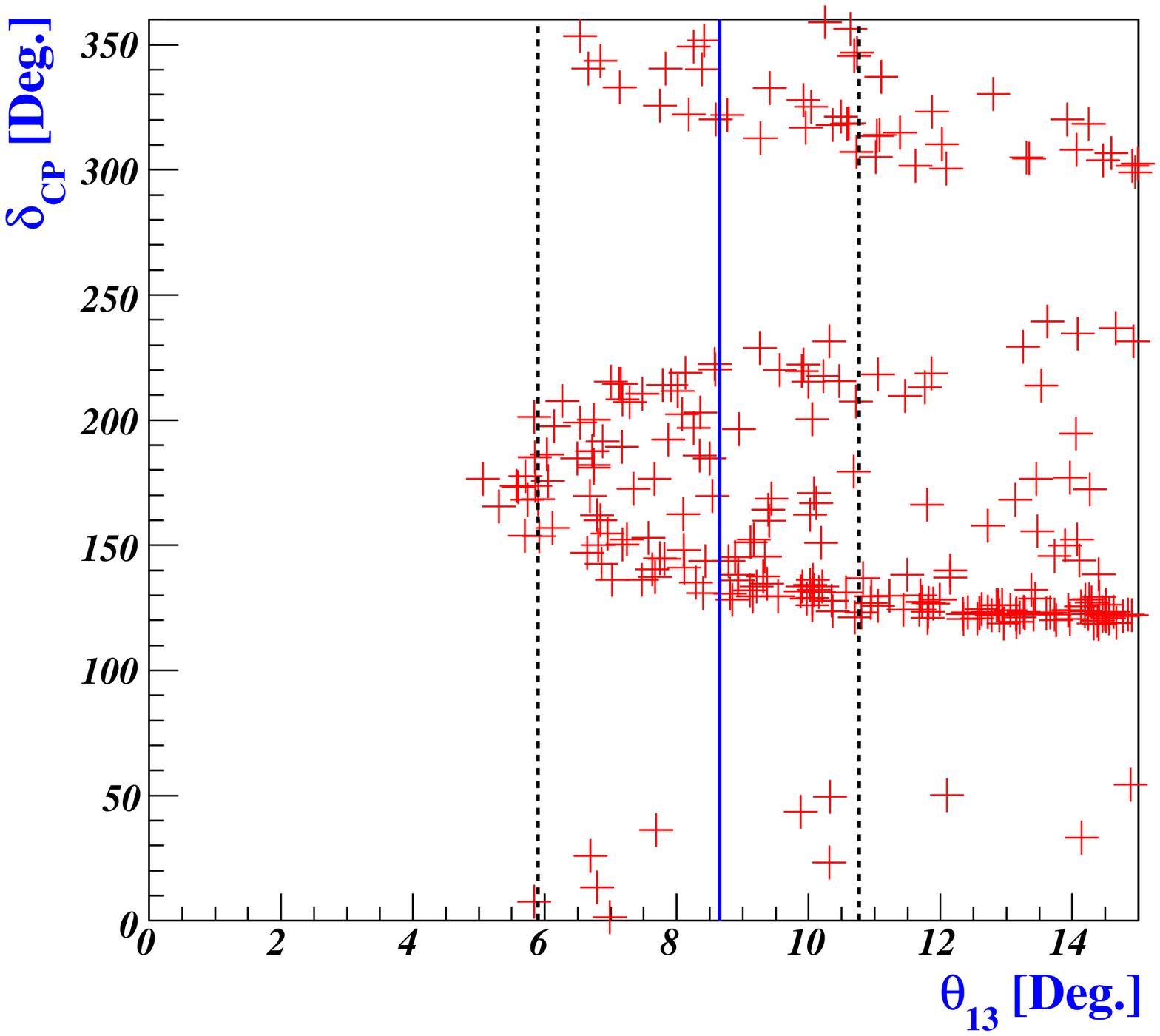,width=6.5cm,angle=0}
\end{minipage}
\hspace*{1.0cm}
\begin{minipage}[t]{6.0cm}
\epsfig{figure=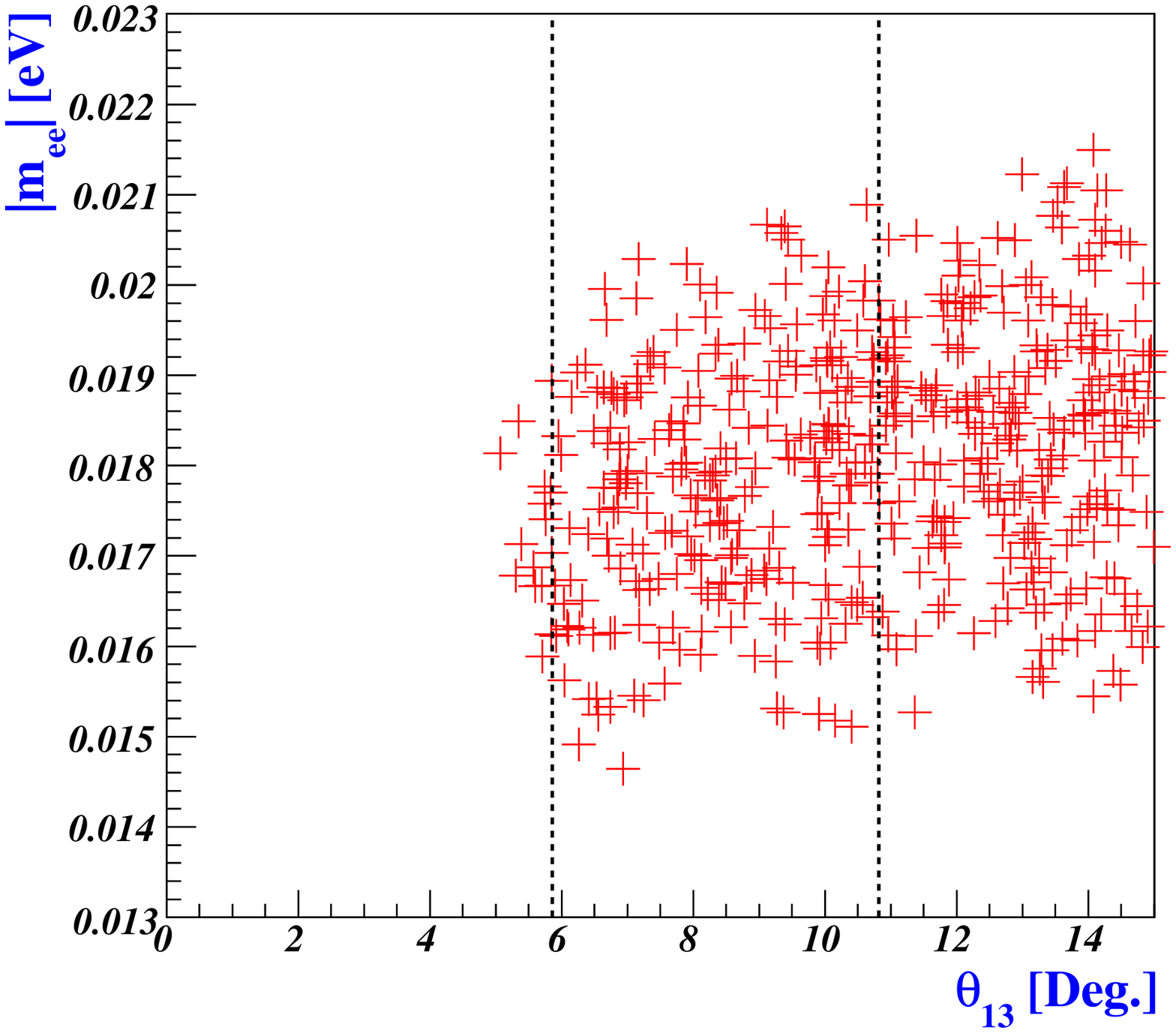,width=6.5cm,angle=0}
\end{minipage}
\caption{\label{FigA2}
 Predictions for the Dirac CP phase $\delta_{CP}$ versus $\theta_{13}$ (left panel) and the effective mass of neutrinoless double beta decay $|m_{ee}|$ versus the mixing angle $\theta_{13}$ (right panel) for Case (i) .
The thick and dotted lines correspond to $\theta_{13}=8.68^{\circ}$ which is the best-fit value and the $3\sigma$ bounds given in Eq.~(\ref{expdata}) including the Daya Bay result, respectively.}
\end{figure}
In the left upper panel of Fig.~\ref{FigA1}, the data points indicate how the mixing angle $\theta_{13}$ is determined in terms of the ratio $y_{1}/y_{2}$.
The result shows that the upper limit of $y_{1}/y_{2}$ is $0.86$, and the measured value of $\theta_{13}$ from the Daya Bay and RENO can be
achieved for two regions, $0.40<y_{1}/y_{2}< 0.57$ and $0.67< y_{1}/y_{2}<0.82$.
To see how the parameters are correlated with low energy {\it CP} violation measurable through
neutrino oscillations, we consider the leptonic {\it CP} violation parameter defined by the
Jarlskog invariant $J_{CP}\equiv{\rm Im}[U_{e1}U_{\mu2}U^{\ast}_{e2}U^{\ast}_{\mu1}]
 =\frac{1}{8}\sin2\theta_{12}\sin2\theta_{23} \sin2\theta_{13}\cos\theta_{13}
 \sin\delta_{CP}$~\cite{Jarlskog:1985ht} which can be described in terms of the elements $h=m_{\nu}m^{\dag}_{\nu}$~\cite{Branco:2002xf}:
 \begin{eqnarray}
  J_{CP}=-\frac{{\rm Im}\{h_{12}h_{23}h_{31}\}}{\Delta m^{2}_{21}\Delta m^{2}_{31}\Delta m^{2}_{32}}~.
  \label{JCP}
 \end{eqnarray}
The behavior of $J_{CP}$ is plotted in the right upper panel of Fig.~\ref{FigA1} as a function of $\theta_{13}$.
We see that the value of $|J_{CP}|$ lies between 0 and $0.034$ for the measured value of $\theta_{13}$.
In our model, since ${\rm Im}\{h_{12}h_{23}h_{31}\}$ is proportional to $1-y^2_2$, the leptonic $CP$ violation $J_{CP}$ goes to zero in the limit of $y_{2}\rightarrow1$.
However, $y_2=1$ is not allowed in our analysis, and thus $J_{CP}=0$ indicates that there exists some cancelation among the terms composed of $\sin\psi_{12}, \sin(\psi_{1}+\psi_{2})$, $\sin(2\psi_{1}-\psi_{2})$ and $\sin\psi_{2}$ multiplies by
$y_{1,2}$, $f(z_{1})/a$, $f(z_{2})$, and $f(z_{3})/b$ even if CP phases $\psi_{1,2}$ are non zero.
In the lower panel of Fig.~\ref{FigA1}, the data points indicate how the values of $\theta_{13}$ depend on $\theta_{12}$ and $\theta_{23}$ in the allowed regions given by Eq.~(\ref{expnu}).
We see that the measured values of $\theta_{13}$ can be achieved for two separate regions of $\theta_{23}$ : $38.6^{\circ}\lesssim\theta_{23}\lesssim43^{\circ}$ and $47^{\circ}\lesssim\theta_{23}\lesssim53.1^{\circ}$, which indicates that the parameter set strongly prefers deviations from maximal mixing for the atmospheric neutrino oscillation.
From the right lower panel of Fig.~\ref{FigA1}, we see that predictions of $\theta_{13}$ does not strongly depend on $\theta_{12}$ for the allowed region.
We see from the figures that $\theta_{13}$ for the normal hierarchy prefers rather large values more than 5 degrees.

We also see from Fig. \ref{FigA1} that  small deviations for $\theta_{23}$ prefer to large value of $\theta_{13}$ in normal hierarchical case.  This can be understood by considering two relations given in Eq.~(21) and Eq.~(26). The phases $\psi_{1,2}$ go to $0$ or $\pi$ as the magnitude of $\kappa$ defined in Eq.(15) decreases, and in the case of $y_{2}=1$ the neutrino mass matrix indicates directly $\theta_{13}=0$ and $\theta_{23}=-\pi/4$. However, deviation of $y_{2}$ from one can be associated with deviation from maximality of atmospheric mixing angle by the following relation,
 \begin{eqnarray}
  \tan\epsilon_{1}=\left(\frac{1+y_{2}}{1-y_{2}}\right)\frac{(1-2y^{2}_{1}+y^{2}_{2})\sin\psi_{1}
  \frac{f(z_{1})f(z_{2})}{a}-(1-y_{2})^{2}\frac{f(z_{3})}{b}\left(\frac{f(z_{1})}{a}\sin\psi_{12}+f(z_{2})\sin\psi_{2}\right)}
  {(1-2y^{2}_{1}+y^{2}_{2})\sin\psi_{1}\frac{f(z_{1})f(z_{2})}{a}-(1+y_{2})^{2}\frac{f(z_{3})}{b}\left(\frac{f(z_{1})}{a}
  \sin\psi_{12}+f(z_{2})\sin\psi_{2}\right)}~.\nonumber
 \end{eqnarray}
This formular for the parameter $\epsilon_1$ is relevant only when $y_{2}\neq1$.
In the case of $y_{2}\rightarrow 1$ while $y_{1}\neq1$ and $\sin\psi_{1}\neq0$, we see from the above equation that the value of $\theta_{23}$ (or $\epsilon_{1}$) can be large but restricted by experimental data. Then, due to Eq~.(26) and $\Omega$ in Eq.~(35),  the value of $\theta_{13}$ gets smaller as $y_{2}\rightarrow 1$.
On the other hand, when $y_{2}$ is much deviated from 1, two cases for $\theta_{23}$ (or $\epsilon_{1}$) are possible.
One is that rather smaller values of $\theta_{23}$ (or $\epsilon_{1}$) are preferred as the value of $\kappa$ (or $\sin\psi_{1}\rightarrow0$ and $\sin\psi_{2}\rightarrow0$) decreases, and the other is that the combination of two parts in numerator of the above equation can lead to wide ranges of $\theta_{23}$ (or $\epsilon_{1}$).
However, when $y_{2}\approx 1$ which makes the above equation irrelevant,
the value of $\theta_{13}$ goes to $0^{\circ}$ (numerically $\lesssim1^{\circ}$), and the value of $\theta_{23}$ can approach $45^{\circ}$ (or $\epsilon_{1}\rightarrow0$) for $y_{1}\rightarrow1$ or $\sin\psi_{1}\rightarrow0$ converge more faster than $y_{2}\rightarrow 1$. We have neglected this case in our paper.

By using the conventional parametrization of the PMNS matrix~\cite{PDG} and Eq.~(\ref{Unu}), one can deduce a expression for Dirac CP phase $\delta_{CP}$ given by
\begin{eqnarray}
  \delta_{CP}
 &=& -\arg \left(\frac{\frac{U^{\ast}_{e1}U_{e3}U_{\tau1}U^{\ast}_{\tau3}}{c_{12}c^{2}_{13}c_{23}s_{13}}+c_{12}c_{23}s_{13}}{s_{12}s_{23}}\right)~.
\label{angle2}
\end{eqnarray}
Moreover, we can straightforwardly obtain the effective neutrino mass $|m_{ee}|$ which is associated with the amplitude for neutrinoless double beta decay :
 \begin{eqnarray}
  |m_{ee}|\equiv|\sum_{i}(U_{\rm PMNS})^{2}_{ei}m_{i}|~,
  \label{mee}
 \end{eqnarray}
where $U_{\rm PMNS}$ is given in Eq.~(\ref{Unu}).
The left panel of Fig.~\ref{FigA2} shows that   $\delta_{CP}$ is predicted to be  $0^{\circ}\lesssim\delta_{CP}\lesssim60^{\circ}$,  $120^{\circ}\lesssim\delta_{CP}\lesssim240^{\circ}$ and $300^{\circ}\lesssim\delta_{CP}\lesssim360^{\circ}$ for the measured values of $\theta_{13}$ at $3\sigma$. In the right panel of Fig.~\ref{FigA2}, we plot the prediction of the effective neutrino mass $|m_{ee}|$ as a function of $\theta_{13}$, which lies between $0.014$ and $0.021$ in the region of the measured values of $\theta_{13}$ at $3\sigma$.

\vskip 0.4cm
{\bf (ii)} inverted hierarchy of light neutrino
\vskip 0.5cm
\begin{figure}[t]
\begin{minipage}[t]{6.0cm}
\epsfig{figure=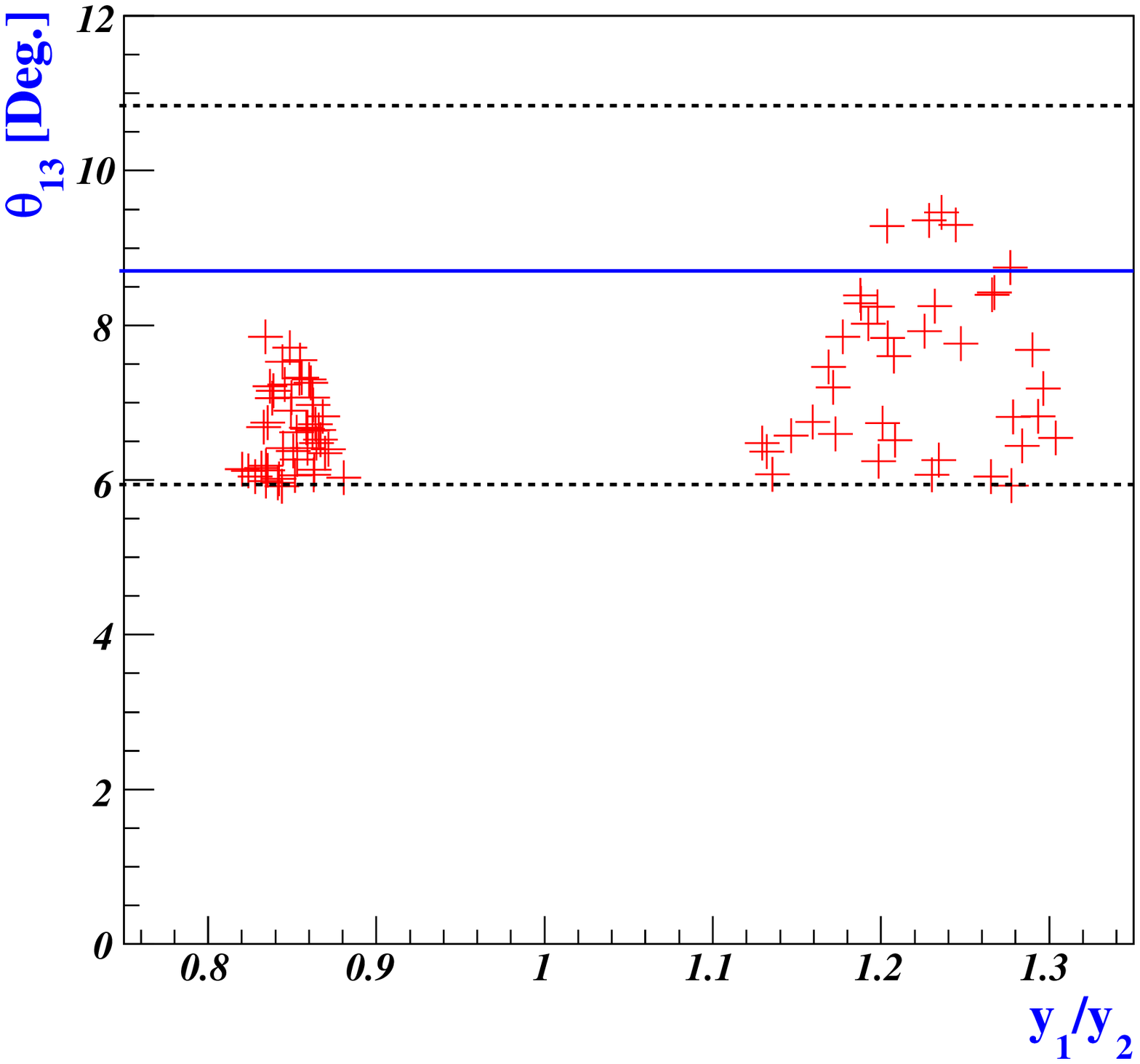,width=6.5cm,angle=0}
\end{minipage}
\hspace*{1.0cm}
\begin{minipage}[t]{6.0cm}
\epsfig{figure=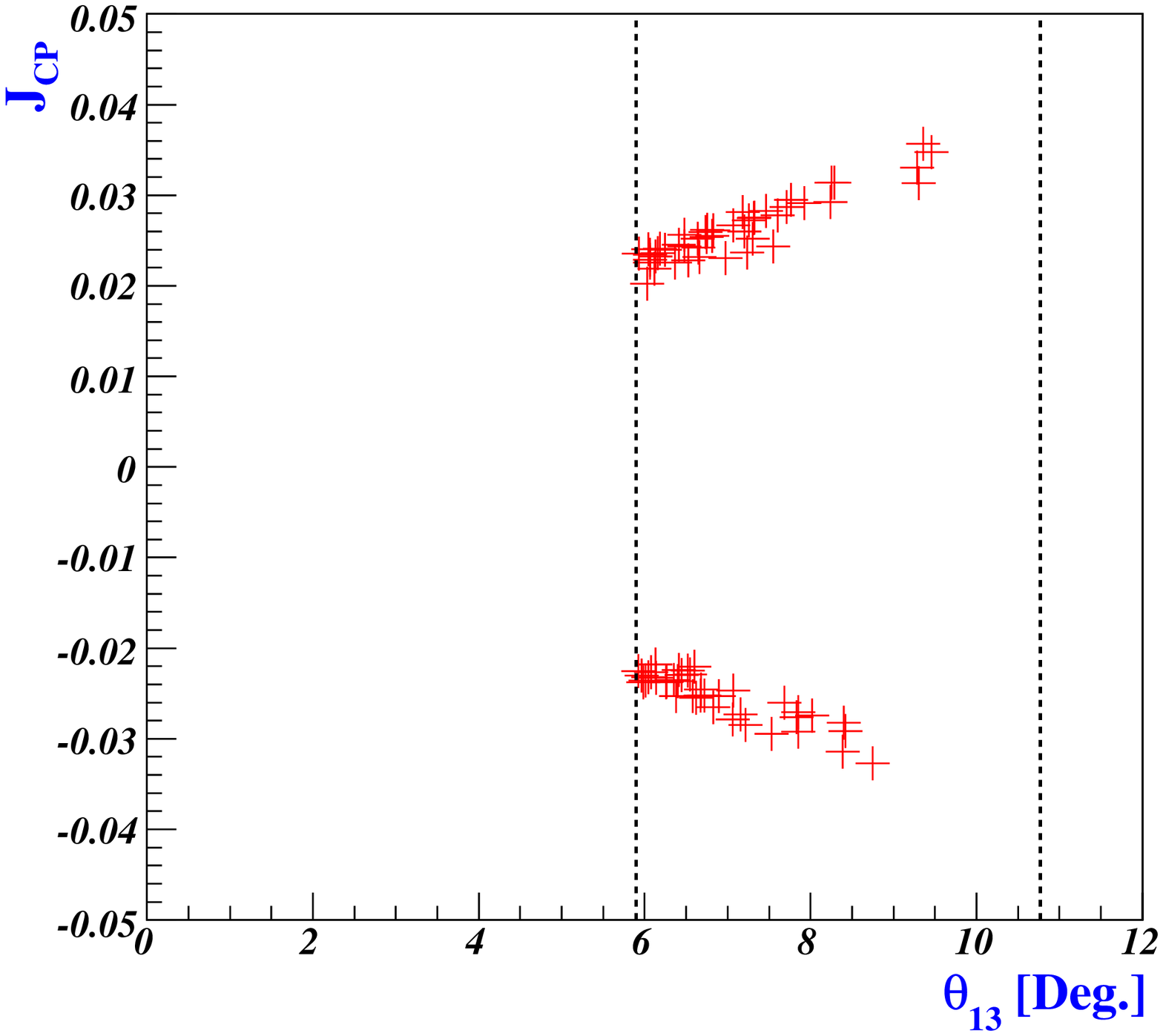,width=6.5cm,angle=0}
\end{minipage}\\
\begin{minipage}[t]{6.0cm}
\epsfig{figure=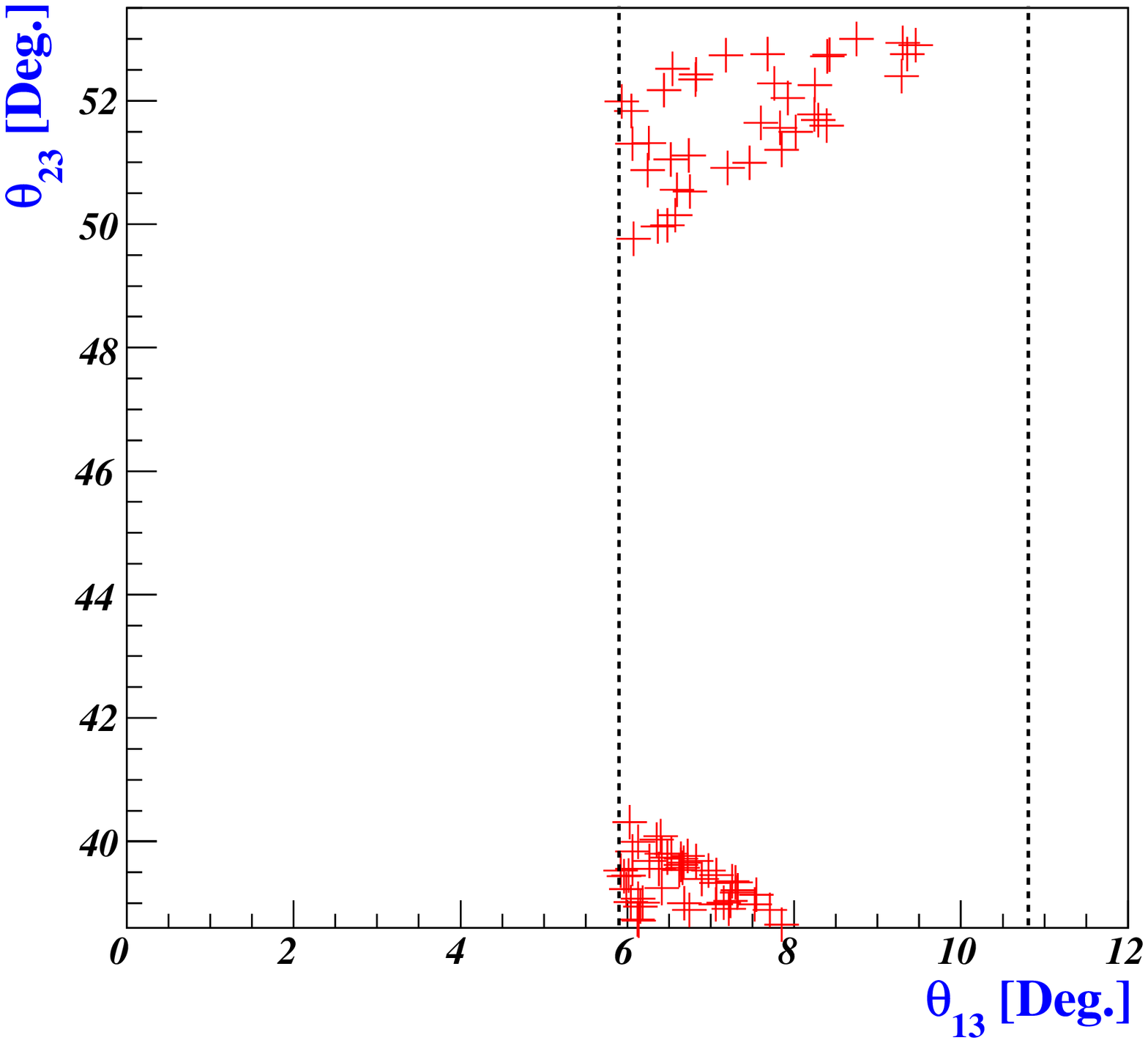,width=6.5cm,angle=0}
\end{minipage}
\hspace*{1.0cm}
\begin{minipage}[t]{6.0cm}
\epsfig{figure=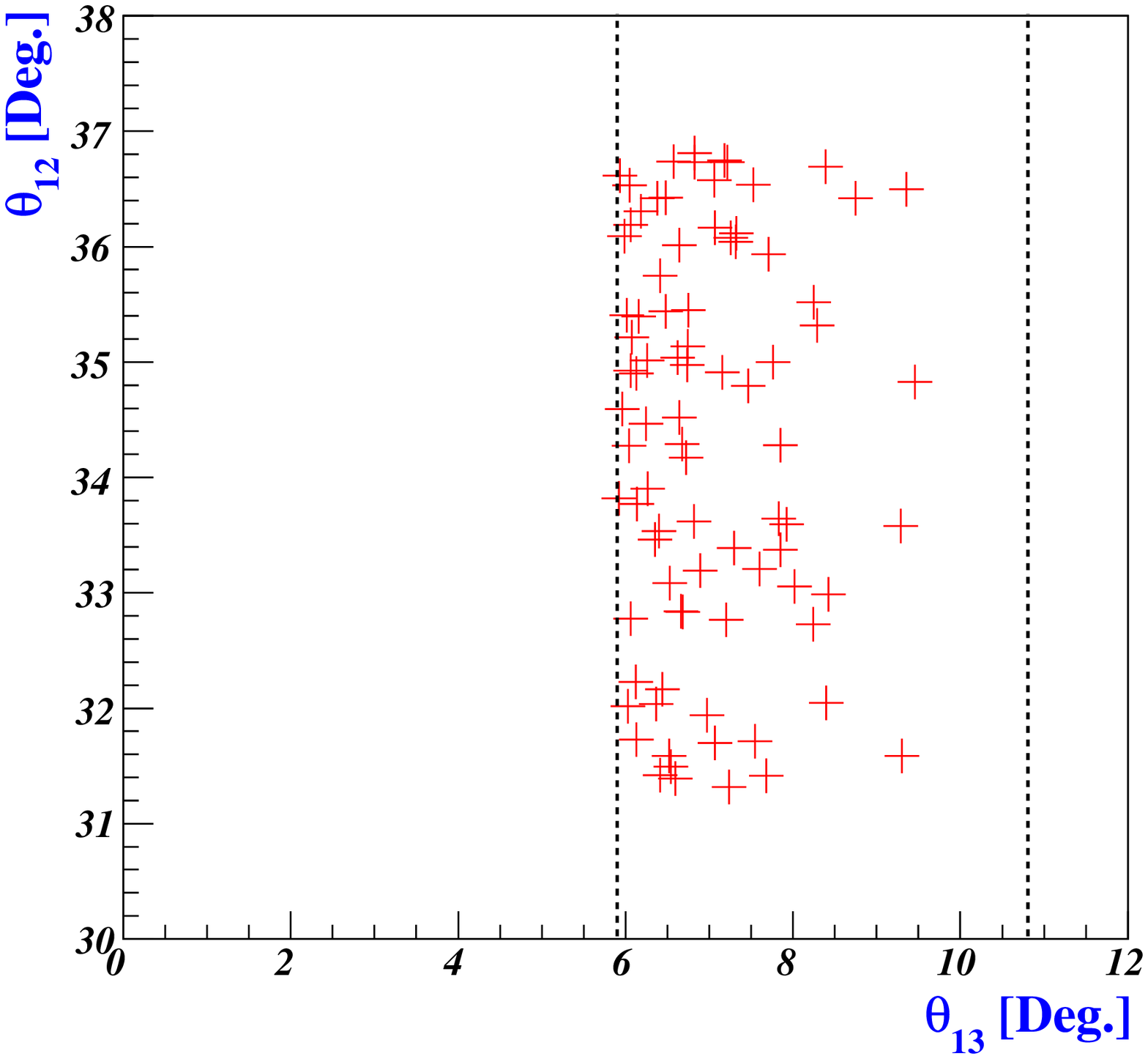,width=6.5cm,angle=0}
\end{minipage}
\caption{\label{FigB1}
Same as Fig.~\ref{FigA1} for Case (ii).}
\end{figure}

Now let us turn to the inverted hierarchical case. Similar to case (i),
scanning the parameters $m_0, y_1, y_2, \kappa, \xi$ based on the formulae for the neutrino mixing angles and masses and
taking the experimental data given at $3\sigma$  in Eq.~(\ref{expnu}) as constraints,  we can obtain the allowed regions of model parameters given by
 \begin{eqnarray}
  &&1.30<\kappa<1.56~,\qquad 209^{\circ}\leq\xi<222^{\circ}~,\qquad 0.27 \leq \frac{y^{\nu}_{3} \lambda_{3}^{\Phi\eta}}{10^{-9}}<0.45~,\nonumber\\
  &&\left\{
      \begin{array}{ll}
        0.79<y_{1}<0.88~,  \\
        0.60<y_{2}<0.79~,
      \end{array}
    \right.~~~{\rm and}~~\left\{
                           \begin{array}{ll}
                             1.12<y_{1}<1.24~, \\
                             1.28<y_{2}<1.5~.
                           \end{array}
                         \right.
  \label{input1}
 \end{eqnarray}
We found that this case is achieved when $M_{1}<M_{2}< M_{3}$ is satisfied.
For those parameter regions, we in turn investigate how the mixing angle $\theta_{13}$ depends on other parameters and whether $CP$ violation is realized.
In the left upper panel of Fig.~\ref{FigB1}, the data points  indicate how the mixing angle $\theta_{13}$ is determined in terms of the ratio $y_{1}/y_{2}$.
We see that  the measured value of $\theta_{13}$ in $3\sigma$ including the Daya Bay experiment in Eq.~(\ref{expdata}),
can be
achieved for two separate regions, $0.82< y_{1}/y_2<0.88$ and $1.12\lesssim y_{1}/y_{2}\lesssim1.3$.
We plot $J_{CP}$ {\it vs.} $\theta_{13}$ in the right upper panel of Fig.~\ref{FigB1}. For $5.9^{\circ}\lesssim\theta_{13}\lesssim 9.5^{\circ}$, $|J_{CP}|\simeq 0.018\sim 0.036$  and $-0.02\sim -0.034$,  which indicates {\it CP} violation
in the leptonic sector.
\begin{figure}[t]
\begin{minipage}[t]{6.0cm}
\epsfig{figure=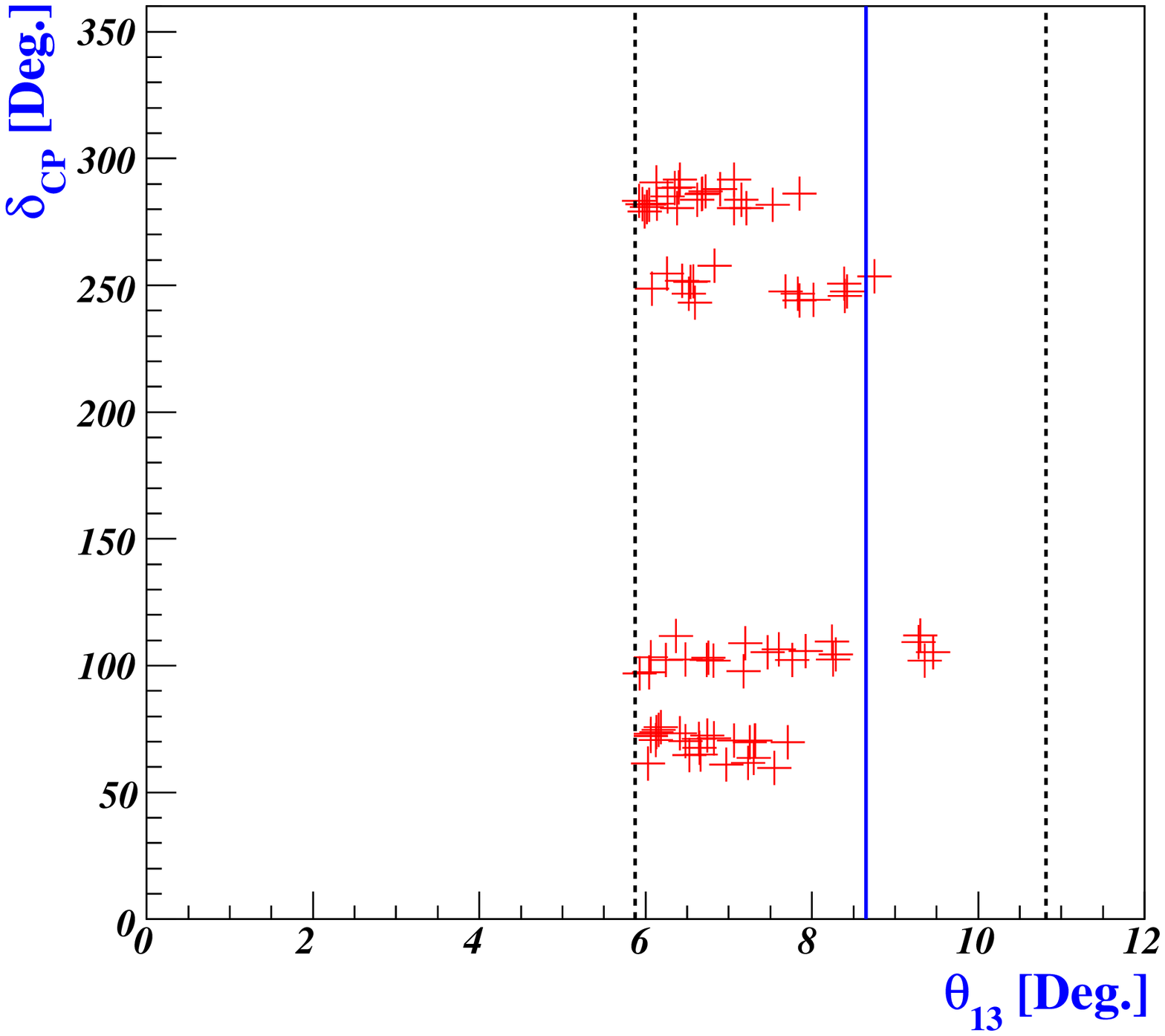,width=6.5cm,angle=0}
\end{minipage}
\hspace*{1.0cm}
\begin{minipage}[t]{6.0cm}
\epsfig{figure=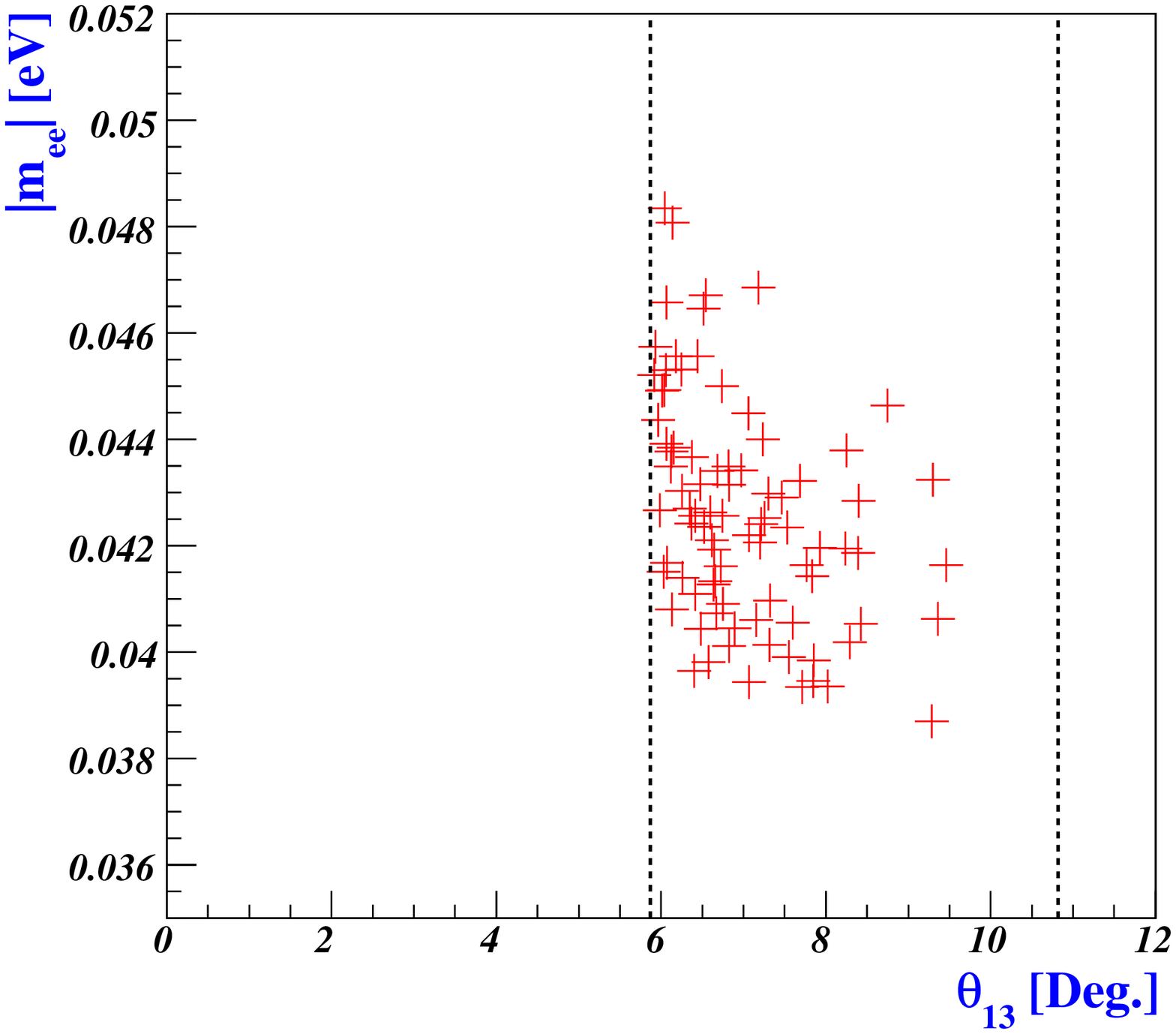,width=6.5cm,angle=0}
\end{minipage}
\caption{\label{FigB2}
Same as Fig.~\ref{FigA2} for Case (ii)}
\end{figure}

In the lower panel of Fig. \ref{FigB1}, the data points show how $\theta_{13}$ is determined in the allowed regions of $\theta_{12}$ and $\theta_{23}$ given by Eq.~(\ref{expnu}).
We see that the narrowed regions of the atmospheric mixing angle $\theta_{23}$, $38.6^{\circ}\lesssim\theta_{23}\lesssim40.5^{\circ}$ and $49.5^{\circ}\lesssim\theta_{23}\lesssim53.1^{\circ}$ are preferred, which indicates
that the parameter set disfavors maximal mixing for the atmospheric mixing angles.
From the lower right panel of Fig.~\ref{FigB1}, we see that determination of $\theta_{13}$ does not strongly depend on $\theta_{12}$ for the allowed region.
We see from the figures that contrary to the case (i), $\theta_{13}$ for the inverted hierarchy prefers rather lower values less than $9.5$ degrees.
The left panel of  Fig.~\ref{FigB2} shows  that  $\delta_{CP}$ is predicted to be around $70^{\circ}$,  $100^{\circ}, 160^{\circ}, 250^{\circ}$ and $290^{\circ}$.  In the right panel of Fig.~\ref{FigB2}, the value of $|m_{ee}|$ is predicted as a function of $\theta_{13}$ and we see that $|m_{ee}|[{\rm eV}]$ lies between  $0.038$ and $0.049$ in the allowed region of
$\theta_{13}$.

\section{Conclusion}
Motivated by recent observations of non-zero $\theta_{13}$ from the Daya Bay and RENO experiments, in this paper, we have proposed a neutrino model with $A_4$ symmetry and
shown how deviations from the TBM mixing indicated by the current neutrino data including
the Daya Bay result can be accounted for.
In addition to the leptons and the Higgs scalar of the SM, our model contains three right handed heavy Majorana neutrinos and several scalar fields
which are electroweak singlets and demanded to construct desirable forms of the letponic mass matrices.
To have a good dark matter candidate, we imposed auxiliary $Z_2$ symmetry, and thus light neutrino masses at tree level are absent in our model.
However, the light neutrino masses can be generated through loop diagram, and we have shown how the light neutrino mass matrix can be diagonalized
by the PMNS mixing matrix whose entries are determined by the current neutrino data including the Daya Bay result.
In our model, the origin of the deviations from TBM mixing is non-degenerate neutrino Yukawa coupling constants among three generations.
Also, unremovable CP phases in the neutrino Yukawa matrix are the origin of the low energy CP violation measurable from neutrino oscillation as well as high energy CP violation.
We have discussed some implication on leptonic CP violation.

\newpage
\appendix


\section{The Higgs mass}
 Our model contains four Higgs doublets and three Higgs singlets.
Here, we present the masses of physical scalar bosons, where the standard Higgs $h'$ is mixed with $\chi'_{0i}$, not with $h'_{i}, A'_{i}$. For simplicity, we assume that CP is conserved in the scalar potential, and then the coupling $\lambda^{\eta\Phi}_{3}$ is real and the term $\xi^{\eta\chi}_{2}(\eta^{\dag}\eta)_{\mathbf{3}_{a}}\chi$ is neglected in the Higgs potential given in Eq.~(\ref{potential}).
The neutral Higgs boson mass matrix in the basis of $(h', \chi'_{01}, \chi'_{02}, \chi'_{03}, h'_{1}, h'_{2}, h'_{3}, A'_{1}, A'_{2}, A'_{3})$ is block diagonalized due to $Z_{2}$ symmetry and CP conservation, which is given by
 \begin{eqnarray}
 {\emph{M}}^{2}_{\rm neutral}=
 {\left(\begin{array}{cccccccccc}
 m^{2}_{h'} & m^{2}_{h'\chi'_1} & 0 & 0 & 0 & 0 & 0 & 0 & 0 & 0 \\
 m^{2}_{h'\chi'_1} & m^{2}_{\chi'_1} & 0 & 0 & 0 & 0 & 0 & 0 & 0 & 0 \\
 0 & 0 & m^{2}_{\chi'_2} & m^{2}_{\chi'_2\chi'_3} & 0 & 0 & 0 & 0 & 0 & 0 \\
 0 & 0 & m^{2}_{\chi'_2\chi'_3} & m^{2}_{\chi'_3} & 0 & 0 & 0 & 0 & 0 & 0 \\
 0 & 0 & 0 & 0 & m^{2}_{h'_1} & 0 & 0 & 0 & 0 & 0 \\
 0 & 0 & 0 & 0 & 0 & m^{2}_{h'_2} & m^{2}_{h'_2h'_3} & 0 & 0 & 0 \\
 0 & 0 & 0 & 0 & 0 & m^{2}_{h'_3h'_2} & m^{2}_{h'_3} & 0 & 0 & 0 \\
 0 & 0 & 0 & 0 & 0 & 0 & 0 & m^{2}_{A'_1} & 0 & 0 \\
 0 & 0 & 0 & 0 & 0 & 0 & 0 & 0 & m^{2}_{A'_2} & m^{2}_{A'_2A'_3} \\
 0 & 0 & 0 & 0 & 0 & 0 & 0 & 0 & m^{2}_{A'_3A'_2} & m^{2}_{A'_3}
 \end{array}\right)}~,
  \label{Higgsmass1}
 \end{eqnarray}
where the primed particles are not mass eigenstates, and mass parameters are given as
 \begin{eqnarray}
  m^{2}_{h'}&=&4\lambda^{\Phi}v^{2}_{\Phi}~,\qquad m^{2}_{h'\chi'_1}=2v_{\Phi}v_{\chi}\lambda^{\Phi\chi}~,\nonumber\\
  m^{2}_{\chi'_1}&=&4v^{2}_{\chi}(\lambda^{\chi}_{1}+\lambda^{\chi}_{2})~,\qquad m^{2}_{\chi'_{2(3)}}=v^{2}_{\chi}(3\lambda^{\chi}_{2}+4\lambda^{\chi}_{3})~,\qquad m^{2}_{\chi'_2\chi'_3}=3v_{\chi}\xi^{\chi}_{1}\nonumber\\
  m^{2}_{h'_1}&=&v^{2}_{\Phi}(\lambda^{\eta\Phi}_{1}+\lambda^{\eta\Phi}_{2}+2\lambda^{\eta\Phi}_{3})+\mu^{2}_{\eta}+v^{2}_{\chi}(\lambda^{\eta\chi}_{1}+2{\rm Re}[\lambda^{\eta\chi}_{2}])~,\nonumber\\
  m^{2}_{A'_1}&=&v^{2}_{\Phi}(\lambda^{\eta\Phi}_{1}+\lambda^{\eta\Phi}_{2}-2\lambda^{\eta\Phi}_{3})+\mu^{2}_{\eta}+v^{2}_{\chi}(\lambda^{\eta\chi}_{1}+2{\rm Re}[\lambda^{\eta\chi}_{2}])~,\nonumber\\
  m^{2}_{h'_{2}}&=&v^{2}_{\Phi}(\lambda^{\eta\Phi}_{1}+\lambda^{\eta\Phi}_{2}+2\lambda^{\eta\Phi}_{3})+\mu^{2}_{\eta}+v^{2}_{\chi}(\lambda^{\eta\chi}_{1}-{\rm Re}[\lambda^{\eta\chi}_{2}]-\sqrt{3}{\rm Im}[\lambda^{\eta\chi}_{2}])~,\nonumber\\
  m^{2}_{h'_{3}}&=&v^{2}_{\Phi}(\lambda^{\eta\Phi}_{1}+\lambda^{\eta\Phi}_{2}+2\lambda^{\eta\Phi}_{3})+\mu^{2}_{\eta}+v^{2}_{\chi}(\lambda^{\eta\chi}_{1}-{\rm Re}[\lambda^{\eta\chi}_{2}]+\sqrt{3}{\rm Im}[\lambda^{\eta\chi}_{2}])~,\nonumber\\
  m^{2}_{A'_{2}}&=&v^{2}_{\Phi}(\lambda^{\eta\Phi}_{1}+\lambda^{\eta\Phi}_{2}-2\lambda^{\eta\Phi}_{3})+\mu^{2}_{\eta}+v^{2}_{\chi}(\lambda^{\eta\chi}_{1}-{\rm Re}[\lambda^{\eta\chi}_{2}]-\sqrt{3}{\rm Im}[\lambda^{\eta\chi}_{2}])~,\nonumber\\
  m^{2}_{A'_{3}}&=&v^{2}_{\Phi}(\lambda^{\eta\Phi}_{1}+\lambda^{\eta\Phi}_{2}-2\lambda^{\eta\Phi}_{3})+\mu^{2}_{\eta}+v^{2}_{\chi}(\lambda^{\eta\chi}_{1}-{\rm Re}[\lambda^{\eta\chi}_{2}]+\sqrt{3}{\rm Im}[\lambda^{\eta\chi}_{2}])~,\nonumber\\
  m^{2}_{h'_2h'_3}&=& m^{2}_{A'_2A'_3}=v_{\chi}\xi^{\eta\chi}_{1}~.
  \label{Higgsmass2}
 \end{eqnarray}
Since the matrix in Eq.~(\ref{Higgsmass1}) is block diagonalized, it is easy to obtain the mass spectrum given as follows;
 \begin{eqnarray}
 m^{2}_{h}&=& \frac{1}{2}\Big\{m^{2}_{h'}+m^{2}_{\chi'_1}-\sqrt{(m^{2}_{h'}-m^{2}_{\chi'_1})^{2}+4(m^{2}_{h'\chi'_1})^2}\Big\}~,\nonumber\\ m^{2}_{\chi_1}&=&  \frac{1}{2}\Big\{m^{2}_{h'}+m^{2}_{\chi'_1}+\sqrt{(m^{2}_{h'}-m^{2}_{\chi'_1})^{2}+4(m^{2}_{h'\chi'_1})^2}\Big\}~,\nonumber\\ m^{2}_{\chi_2}&=& m^{2}_{\chi'_2}-m^{2}_{\chi'_2\chi'_3}~,\quad m^{2}_{\chi_3}= m^{2}_{\chi'_2}+m^{2}_{\chi'_2\chi'_3}~,\nonumber\\
 m^{2}_{h_1}&=&m^{2}_{h'_1}~,\qquad m^{2}_{A_1}=m^{2}_{A'_1} ~,\nonumber\\
 m^{2}_{h_2}&=& v^{2}_{\Phi}(\lambda^{\eta\Phi}_{12}+2\lambda^{\eta\Phi}_{3})+\mu^{2}_{\eta}+v^{2}_{\chi}(\lambda^{\eta\chi}_{1}-{\rm Re}[\lambda^{\eta\chi}_{2}])-v_{\chi}\sqrt{3(v_{\chi}{\rm Re}[\lambda^{\eta\Phi}_{2}])^{2}+(\xi^{\eta\chi}_{1})^{2}}~,\nonumber\\
 m^{2}_{A_2}&=&  v^{2}_{\Phi}(\lambda^{\eta\Phi}_{12}-2\lambda^{\eta\Phi}_{3})+\mu^{2}_{\eta}+v^{2}_{\chi}(\lambda^{\eta\chi}_{1}-{\rm Re}[\lambda^{\eta\chi}_{2}])-v_{\chi}\sqrt{3(v_{\chi}{\rm Re}[\lambda^{\eta\Phi}_{2}])^{2}+(\xi^{\eta\chi}_{1})^{2}}~,\nonumber\\
 m^{2}_{h_3}&=&  v^{2}_{\Phi}(\lambda^{\eta\Phi}_{12}+2\lambda^{\eta\Phi}_{3})+\mu^{2}_{\eta}+v^{2}_{\chi}(\lambda^{\eta\chi}_{1}-{\rm Re}[\lambda^{\eta\chi}_{2}])+v_{\chi}\sqrt{3(v_{\chi}{\rm Re}[\lambda^{\eta\Phi}_{2}])^{2}+(\xi^{\eta\chi}_{1})^{2}}~,\nonumber\\
 m^{2}_{A_3}&=&  v^{2}_{\Phi}(\lambda^{\eta\Phi}_{12}-2\lambda^{\eta\Phi}_{3})+\mu^{2}_{\eta}+v^{2}_{\chi}(\lambda^{\eta\chi}_{1}-{\rm Re}[\lambda^{\eta\chi}_{2}])+v_{\chi}\sqrt{3(v_{\chi}{\rm Re}[\lambda^{\eta\Phi}_{2}])^{2}+(\xi^{\eta\chi}_{1})^{2}}~,
 \end{eqnarray}
 where $\lambda^{\eta\Phi}_{12}\equiv \lambda^{\eta\Phi}_{1} +\lambda^{\eta\Phi}_{2}$.
Note here that the unprimed particles denote mass eigenstates.
And the charged Higgs boson mass matrix in the basis of $(\eta^{\pm}_{1} ,\eta^{\pm}_{2}, \eta^{\pm}_{3})$ is given as
 \begin{eqnarray}
 m^{2}_{\rm charged}=
 {\left(\begin{array}{ccc}
 m^{2}_{\eta^{\pm}_{1}} &  0 & 0  \\
 0 & m^{2}_{\eta^{\pm}_{2}} & 0  \\
 0 & 0 & m^{2}_{\eta^{\pm}_{3}}
 \end{array}\right)}~,
  \label{Higgsmass3}
 \end{eqnarray}
where
 \begin{eqnarray}
  m^{2}_{\eta^{\pm}_{1}}&=&\mu^{2}_{\eta}+v^{2}_{\Phi}\lambda^{\eta\Phi}_{1}+v^{2}_{\chi}\left(\lambda^{\eta\chi}_{1}+2{\rm Re}[\lambda^{\eta\chi}_{2}]\right)~,\nonumber\\
  m^{2}_{\eta^{\pm}_{2}}&=&\mu^{2}_{\eta}+v^{2}_{\Phi}\lambda^{\eta\Phi}_{1}+v^{2}_{\chi}\left(\lambda^{\eta\chi}_{1}-{\rm Re}[\lambda^{\eta\chi}_{2}]-\sqrt{3}{\rm Im}[\lambda^{\eta\chi}_{2}]\right)~,\nonumber\\
  m^{2}_{\eta^{\pm}_{3}}&=&\mu^{2}_{\eta}+v^{2}_{\Phi}\lambda^{\eta\Phi}_{1}+v^{2}_{\chi}\left(\lambda^{\eta\chi}_{1}-{\rm Re}[\lambda^{\eta\chi}_{2}]+\sqrt{3}{\rm Im}[\lambda^{\eta\chi}_{2}]\right)~.
 \end{eqnarray}

Using $m^{2}_{h_i}$, $m^{2}_{A_i}$ in Eq.~(\ref{Higgsmass2}) and Eq.~(\ref{Higgsmass3}), the expressions for $\bar{m}^{2}_{\eta_{i}}$ appeared in Eq.~(\ref{lownu1}) are
 \begin{eqnarray}
  \bar{m}^{2}_{\eta_{1}}&=&\mu^{2}_{\eta}+v^{2}_{\Phi}\lambda^{\eta\Phi}_{12}+v^{2}_{\chi}\left(\lambda^{\eta\chi}_{1}+2{\rm Re}[\lambda^{\eta\chi}_{2}]\right)
  =m^{2}_{\eta^{\pm}_{1}}+v^{2}_{\Phi}\lambda^{\eta\Phi}_{2}~,\nonumber\\
  \bar{m}^{2}_{\eta_{2}}&=&v^{2}_{\Phi}\lambda^{\eta\Phi}_{12}+\mu^{2}_{\eta}+v^{2}_{\chi}(\lambda^{\eta\chi}_{1}-{\rm Re}[\lambda^{\eta\chi}_{2}]-\sqrt{3}{\rm Im}[\lambda^{\eta\chi}_{2}])=m^{2}_{\eta^{\pm}_{2}}+v^{2}_{\Phi}\lambda^{\eta\Phi}_{2}~,\nonumber\\
  \bar{m}^{2}_{\eta_{3}}&=&v^{2}_{\Phi}\lambda^{\eta\Phi}_{12}+\mu^{2}_{\eta}+v^{2}_{\chi}(\lambda^{\eta\chi}_{1}-{\rm Re}[\lambda^{\eta\chi}_{2}]+\sqrt{3}{\rm Im}[\lambda^{\eta\chi}_{2}])=m^{2}_{\eta^{\pm}_{3}}+v^{2}_{\Phi}\lambda^{\eta\Phi}_{2}~,
  \label{Higgsmass4}
 \end{eqnarray}

\acknowledgments{The work of S.K. Kang was supported in part by the National Research Foundation of Korea (NRF) grant funded by the Korea government of the Ministry of Education, Science and Technology (MEST) (No. 2011-0003287).
}


\end{document}